\documentclass[10pt]{article}

\usepackage{mathptmx,amsmath,amsthm,amsfonts,amssymb,graphicx}

\usepackage[margin=1in,footskip=.5in]{geometry}

\newtheorem{theorem}{Theorem}[section]
\newtheorem{conjecture}[theorem]{Conjecture}
\newtheorem{corollary}[theorem]{Corollary}
\newtheorem{definition}[theorem]{Definition}
\newtheorem{lemma}[theorem]{Lemma}

\newtheorem{proposition}[theorem]{Proposition}

\begin{document}

\title{Quantum versus classical proofs and advice}

\author{Scott Aaronson, MIT (currently UT Austin)
    \thanks{Email: aaronson@cs.utexas.edu}
    \and Greg Kuperberg, UC Davis \thanks{Email: greg@math.ucdavis.edu}}

\maketitle

\begin{abstract}
This paper studies whether quantum proofs are more powerful than
classical proofs, or in complexity terms, whether
$\mathsf{QMA}=\mathsf{QCMA}$. \ We prove three results about this
question. \ First, we give a \textquotedblleft quantum oracle
separation\textquotedblright\ between $\mathsf{QMA}$ and
$\mathsf{QCMA}$. \ More concretely, we show that any quantum
algorithm needs $\Omega\left(  \sqrt{\frac{2^{n}}{m+1}}\right)  $
queries to find an $n$-qubit \textquotedblleft marked
state\textquotedblright\ $\left\vert \psi \right\rangle $,\ even if
given an $m$-bit classical description of $\left\vert
\psi\right\rangle $ together with a quantum black box that
recognizes $\left\vert \psi\right\rangle $. \ Second, we give an
explicit $\mathsf{QCMA}$ protocol that nearly achieves this lower
bound. \ Third, we show that, in the one previously-known case where
quantum proofs seemed to provide an exponential advantage,
\textit{classical} proofs are basically just as powerful. \ In
particular, Watrous gave a $\mathsf{QMA}$\ protocol for verifying
non-membership in finite groups. \ Under plausible group-theoretic
assumptions, we give a $\mathsf{QCMA}$\ protocol for the same
problem. \ Even with no assumptions, our protocol makes only
polynomially many queries to the group oracle. \ We end with some
conjectures about quantum versus classical oracles, and about the
possibility of a \textit{classical} oracle separation between
$\mathsf{QMA}$\ and $\mathsf{QCMA}$.
\end{abstract}

\section{Introduction\label{INTRO}}

If someone hands you a quantum state, is that more \textquotedblleft
useful\textquotedblright\ than being handed a classical string with a
comparable number of bits? \ In particular, are there truths that you can
efficiently verify, and are there problems that you can efficiently solve,
using the quantum state but not using the string? \ These are the questions
that this paper addresses, and that it answers in several contexts.

Recall that $\mathsf{QMA}$, or Quantum Merlin-Arthur, is the class of decision
problems for which a \textquotedblleft yes\textquotedblright\ answer can be
verified in quantum polynomial time, with help from a polynomial-size quantum
witness state $\left\vert \psi\right\rangle $. \ Many results are known about
$\mathsf{QMA}$: for example, it has natural complete promise problems
\cite{kkr}, allows amplification of success probabilities \cite{mw}, and is
contained in $\mathsf{PP}$ \cite{mw}. \ Raz and Shpilka \cite{razshpilka}%
\ have also studied communication complexity variants of $\mathsf{QMA}$.

Yet as Aharonov and Naveh \cite{an}\ pointed out in 2002, the very definition
of $\mathsf{QMA}$\ raises a fundamental question. \ Namely: is it really
essential that the witness be quantum, or does it suffice for the\textit{
}algorithm \textit{verifying }the witness to be quantum? \ To address this
question, Aharonov and Naveh\ defined the class $\mathsf{QCMA}$, or
\textquotedblleft Quantum Classical Merlin-Arthur,\textquotedblright\ to be
the same as $\mathsf{QMA}$\ except that now the witness is
classical.\footnote{Some say that this class would more accurately be called
$\mathsf{CMQA}$, for \textquotedblleft Classical Merlin Quantum
Arthur.\textquotedblright\ \ But $\mathsf{QCMA}$\ has stuck.} \ We can then
ask whether $\mathsf{QMA}=\mathsf{QCMA}$. \ Not surprisingly, the answer is
that we don't know.

If we can't decide whether two complexity classes are equal, the usual next
step is to construct a relativized world that separates them. \ This would
provide at least some evidence that the classes are different. \ But in the
case of $\mathsf{QMA}$\ versus $\mathsf{QCMA}$, even this limited goal has
remained elusive.

Closely related to the question of quantum versus classical proofs is that of
quantum versus classical \textit{advice}. \ Compared to a proof, advice has
the advantage that it can be trusted, but the disadvantage that it can't be
tailored to a particular input. \ More formally, let $\mathsf{BQP/qpoly}$\ be
the class of problems solvable in quantum polynomial time, with help from a
polynomial-size \textquotedblleft quantum advice state\textquotedblright%
\ $\left\vert \psi_{n}\right\rangle $\ that depends only on the input length
$n$. \ Then the question is whether $\mathsf{BQP/qpoly}=\mathsf{BQP/poly}$,
where $\mathsf{BQP/poly}$\ is the class of problems solvable in quantum
polynomial time with help from polynomial-size \textit{classical} advice.
\ Aaronson \cite{aar:adv}\ showed that $\mathsf{BQP/qpoly}\subseteq
\mathsf{PP/poly}$, which at least tells us that quantum advice is not
\textquotedblleft infinitely\textquotedblright\ more powerful than classical
advice. \ But, like the $\mathsf{QMA}$ versus $\mathsf{QCMA}$ question, the
$\mathsf{BQP/qpoly}$ versus $\mathsf{BQP/poly}$\ question has remained open,
with not even an oracle separation known.

\subsection{Our Results\label{RESULTS}}

This paper introduces new tools with which to attack $\mathsf{QMA}$ versus
$\mathsf{QCMA}$\ and related questions.

First, we achieve an oracle separation between $\mathsf{QMA}$ and
$\mathsf{QCMA}$, but only by broadening the definition of \textquotedblleft
oracle.\textquotedblright\ \ In particular, we introduce the notion of a
\textit{quantum oracle}, which is just an infinite sequence of unitaries
$U=\left\{  U_{n}\right\}  _{n\geq1}$\ that a quantum algorithm can apply in a
black-box fashion. \ Just as a classical oracle models a subroutine to which
an algorithm has black-box access, so a quantum oracle models a quantum
subroutine, which can take quantum input and produce quantum output. \ We are
able to give a quantum oracle\ that separates $\mathsf{QMA}$ from
$\mathsf{QCMA}$:

\begin{theorem}
\label{qosepthm}There exists a quantum oracle $U$ such that $\mathsf{QMA}%
^{U}\neq\mathsf{QCMA}^{U}$.
\end{theorem}

Similarly, there exists a quantum oracle $V$ such that $\mathsf{BQP}%
^{V}\mathsf{/qpoly}\neq\mathsf{BQP}^{V}\mathsf{/poly}$.

Theorem \ref{qosepthm}\ implies that if $\mathsf{QMA}=\mathsf{QCMA}$, then any
proof of this fact will require \textquotedblleft quantumly nonrelativizing
techniques\textquotedblright: techniques that are sensitive to the presence of
quantum oracles. \ Currently, we do not know of \textit{any} quantumly
nonrelativizing techniques that are not also classically nonrelativizing.
\ For this reason, we believe that quantum oracle separations merit the same
informal interpretation as classical oracle separations: almost any argument
that one might advance against the former, is also an argument against the
latter! \ The difference is that quantum oracle results are sometimes much
easier to prove than classical ones. \ To our knowledge, this paper provides
the first example of this phenomenon, but other examples have since emerged
\cite{aar:copy,moscastebila}.

It might be objected that, even if quantum oracle separations are no less
trustworthy than classical ones, they certainly aren't \textit{more}
trustworthy, and complexity theorists have known since the celebrated
$\mathsf{IP}=\mathsf{PSPACE}$\ theorem \cite{shamir}\ that oracle results
sometimes \textquotedblleft point in the wrong direction.\textquotedblright%
\ \ We wish to stress two points in response. \ First, oracle results provide
at least \textit{some} understanding, thereby opening the way to further
progress. \ This is particularly true in quantum computing, where even the
oracle results tend to be much less intuitively obvious than they are in the
classical world. \ Second, complexity theorists do not currently have any
nonrelativizing technique for \textquotedblleft
non-interactive\textquotedblright\ classes such as $\mathsf{QMA}$ and
$\mathsf{QCMA}$ even remotely analogous to the arithmetization technique that
Shamir \cite{shamir}\ used to show $\mathsf{IP}=\mathsf{PSPACE}$. \ We hope
such a technique will someday be discovered.

Underlying Theorem \ref{qosepthm}\ is the following lower bound. \ Suppose a
unitary oracle $U_{n}$\ acts on $n$ qubits, and suppose that either (i)
$U_{n}$ is the identity, or (ii) there exists a secret $n$-qubit
\textquotedblleft marked state\textquotedblright\ $\left\vert \psi
_{n}\right\rangle $\ such that $U_{n}\left\vert \psi_{n}\right\rangle
=-\left\vert \psi_{n}\right\rangle $, but $U_{n}\left\vert \varphi
\right\rangle =\left\vert \varphi\right\rangle $\ whenever $\left\vert
\varphi\right\rangle $\ is orthogonal to $\left\vert \psi_{n}\right\rangle $.
\ Then even if a quantum algorithm is given $m$ bits of classical advice about
$\left\vert \psi_{n}\right\rangle $, the algorithm still needs $\Omega\left(
\sqrt{\frac{2^{n}}{m+1}}\right)  $\ queries to $U_{n}$\ to distinguish these
cases. \ Note that when $m=0$, we recover the usual $\Omega\left(  \sqrt
{2^{n}}\right)  $\ lower bound for Grover search as a special case. \ At the
other extreme, if $m\approx2^{n}$ then our bound gives nothing---not
surprisingly, since the classical advice might contain explicit instructions
for preparing $\left\vert \psi_{n}\right\rangle $. \ The point is that, if $m$
is \textit{not} exponentially large, then exponentially many queries are needed.

Since $\left\vert \psi_{n}\right\rangle $\ is an arbitrary $2^{n}$-dimensional
unit vector, it might be thought obvious\ that $2^{\Omega\left(  n\right)}$\
bits are needed to describe that vector. \ The key point, however, is that
the\ $\mathsf{QCMA}$\ verifier is given not only a classical description of
$\left\vert \psi_{n}\right\rangle $, but also oracle access to $U_{n}$. \ So
the question is whether some \textit{combination} of these resources might be
exponentially more powerful than either one alone. \ We prove that the answer
is no, using the hybrid argument of Bennett et al. \cite{bbbv} together
with geometric results about partitioning the unit sphere.

In Section \ref{UPPER}, we show that our lower bound is basically tight, by
giving an algorithm that finds $\left\vert \psi_{n}\right\rangle $\ using
$O\left(  \sqrt{2^{n}/m}\right)  $\ queries when $m\geq2n$. \ This algorithm
has the drawback of being \textit{computationally} inefficient. \ To fix this,
we give another algorithm that finds $\left\vert \psi_{n}\right\rangle
$\ using\ $O\left(  n\sqrt{2^{n}/m}\right)  $\ queries together with $O\left(
n^{2}\sqrt{2^{n}/m}+\operatorname*{poly}\left(  m\right)  \right)
$\ computational steps.

Having separated $\mathsf{QMA}$\ from $\mathsf{QCMA}$ by a quantum
oracle,\ we next revisit the question of whether these classes can
be separated by a \textit{classical} oracle. \ Right now, we know of
only one candidate problem for such a separation in the literature:
the Group Non-Membership ($\operatorname*{GNM}$) problem, which
Watrous \cite{watrous} placed in $\mathsf{QMA}$ even though Babai
\cite{babai:am} showed it is not in $\mathsf{MA}$ as an oracle
problem.\footnote{Interestingly, the classes $\mathsf{MA}$ and
$\mathsf{AM}$ were originally defined by Babai in connection with
GNM \cite{babai:am2}.} In Group Non-Membership, Arthur is given
black-box access to a finite group $G$, together with a subgroup
$H\leq G$ specified by its generators and an element $x\in G$. \
Arthur's goal is to verify that $x\notin H$, using a number of group
operations polynomial in $\log\left\vert G\right\vert $. \ (Note
that the group \textit{membership} problem is in $\mathsf{NP}$ by a
result of Babai and Szemer\'{e}di \cite{bs:matrix}.) \ In Watrous's
protocol, the quantum witness is simply an equal superposition
$\left\vert H\right\rangle $\ over the elements of $H$. \ Given such
a witness, Arthur can check non-membership by comparing the states
$\left\vert H\right\rangle $ and $\left\vert xH\right\rangle $, and
can similarly check the veracity of $\left\vert H\right\rangle $ by
comparing it to $\left\vert hH\right\rangle $, where $h$ is an
almost-uniformly random element of $H$.

Evidently a classical proof of non-membership would have to be completely
different. \ Nevertheless, in Section \ref{GNM}\ we show the following:

\begin{theorem}
\label{hspthm}$\operatorname*{GNM}$ has polynomially-bounded $\mathsf{QCMA}$
query complexity.
\end{theorem}

Theorem \ref{hspthm}\ implies that it is pointless to try to prove a classical
oracle separation between $\mathsf{QMA}$\ and $\mathsf{QCMA}$ by proving a
lower bound on the quantum query complexity of Group Non-Membership. \ If such
a separation is possible, then a new approach will be needed.

The idea of the proof of Theorem \ref{hspthm} is that Merlin can
\textquotedblleft pull the group out of the black box.\textquotedblright\ \ In
other words, he can claim an embedding of a model group $\Gamma$ into $G$.
\ This claim is entirely classical, but verifying it requires solving the
Normal Hidden Subgroup Problem ($\operatorname*{NHSP}$) in $\Gamma$. \ This
problem has low query complexity by a result of Ettinger, H\o yer, and Knill
\cite{ehk}, but is not known to be in $\mathsf{BQP}$. \ In addition, analyzing
the description of $\Gamma$ is not known to be computationally efficient.
\ Nonetheless, in Section \ref{COMP} we discuss evidence that
$\operatorname*{NHSP}$ is in $\mathsf{BQP}$ and that non-membership for
$\Gamma$ is in $\mathsf{NP}$. \ Based on this evidence, we conjecture the
following:

\begin{conjecture}
\label{gnminqcma}$\operatorname*{GNM}$ is in $\mathsf{QCMA}$.
\end{conjecture}

Given our results in Section \ref{GNM}, the question remains of whether there
is some other way to prove a classical oracle separation between
$\mathsf{QMA}$\ and $\mathsf{QCMA}$. \ In Section \ref{SQO}, we conjecture
that the answer is yes:

\begin{conjecture}
\label{sepconj}There exists a classical oracle $A$ such that $\mathsf{QMA}%
^{A}\neq\mathsf{QCMA}^{A}$. \ Furthermore, this can be proven by exhibiting an
oracle problem with polynomial $\mathsf{QMA}$\ query complexity but
exponential $\mathsf{QCMA}$\ query complexity.
\end{conjecture}

The reason we believe Conjecture \ref{sepconj}\ is that it seems possible, for
many purposes, to \textquotedblleft encode\textquotedblright\ a quantum oracle
into a classical one. \ In Section \ref{SQO} we explain more concretely what
we mean by that, and present some preliminary results. \ For example, we show
that there exists a $\mathsf{BQP}$ algorithm that maps an oracle string $A$ to
an $n$-qubit pure state $\left\vert \psi_{A}\right\rangle $, such that if
$A$\ is uniformly random, then $\left\vert \psi_{A}\right\rangle $\ is (under
a suitable metric) close to uniformly random under the Haar measure. \ We also
study the question of applying a random $N\times N$ unitary matrix using a
random classical oracle in the same way. \ We do not know how to do this, but
we show that one quantum query will not suffice for this purpose. \ To prove
this, we show that a quantum algorithm that uses just one query can apply at
most $4^{N}$ different $N\times N$\ unitaries, whereas the number of unitaries
required to approximate the uniform distribution grows like $2^{\Theta\left(
N^{2}\right)  }$.

We end in Section \ref{OPEN}\ with some open problems.

\section{Preliminaries\label{PRELIM}}

Throughout this paper, we refer to the set of $N$-dimensional pure states as
$\mathbb{CP}^{N-1}$\ (that is, complex projective space with $N-1$%
\ dimensions). \ We use $\Pr$\ to denote probability, and $\operatorname*{E}%
$\ to denote expectation.

We assume familiarity with standard complexity classes such as $\mathsf{BQP}%
$\ and $\mathsf{MA}$. \ For completeness, we now define $\mathsf{QMA}%
$,\ $\mathsf{QCMA}$, $\mathsf{BQP/qpoly}$, and $\mathsf{BQP/poly}$.

\begin{definition}
$\mathsf{QMA}$ is the class of languages $L\subseteq\left\{  0,1\right\}
^{n}$\ for which there exists a polynomial-time quantum verifier $\mathcal{Q}$
and a polynomial $p$ such that, for all $x\in\left\{  0,1\right\}  ^{n}$:

\begin{enumerate}
\item[(i)] If $x\in L$\ then there exists a $p\left(  n\right)  $-qubit
quantum proof $\left\vert \varphi\right\rangle $\ such that $\mathcal{Q}%
$\ accepts with probability at least $2/3$\ given $\left\vert x\right\rangle
\left\vert \varphi\right\rangle $\ as input.

\item[(ii)] If $x\notin L$\ then $\mathcal{Q}$\ accepts with probability at
most $1/3$\ given $\left\vert x\right\rangle \left\vert \varphi\right\rangle
$\ as input,\ for all purported proofs $\left\vert \varphi\right\rangle $.
\end{enumerate}

The class $\mathsf{QCMA}$ is defined similarly, except that $\left\vert
\varphi\right\rangle $\ is replaced by a\ classical string $z\in\left\{
0,1\right\}  ^{p\left(  n\right)  }$.
\end{definition}

\begin{definition}
$\mathsf{BQP/qpoly}$ is the class of languages $L\subseteq\left\{
0,1\right\}  ^{n}$\ for which there exists a polynomial-time quantum algorithm
$\mathcal{Q}$, together with a set of states $\left\{  \left\vert \psi
_{n}\right\rangle \right\}  _{n\geq1}$\ (where $\left\vert \psi_{n}%
\right\rangle $\ has size $p\left(  n\right)  $\ for some polynomial $p$),
such that for all $x\in\left\{  0,1\right\}  ^{n}$:

\begin{enumerate}
\item[(i)] If $x\in L$\ then $\mathcal{Q}$\ accepts with probability at least
$2/3$\ given $\left\vert x\right\rangle \left\vert \psi_{n}\right\rangle $\ as
input.

\item[(ii)] If $x\notin L$\ then $\mathcal{Q}$\ accepts with probability at
most $1/3$\ given $\left\vert x\right\rangle \left\vert \psi_{n}\right\rangle
$\ as input.
\end{enumerate}

The class $\mathsf{BQP/poly}$ is defined similarly, except that $\left\vert
\psi_{n}\right\rangle $\ is replaced by a\ classical string $a_{n}\in\left\{
0,1\right\}  ^{p\left(  n\right)  }$.
\end{definition}

Let us now explain what we mean by a \textquotedblleft quantum
oracle.\textquotedblright\ \ For us, a quantum oracle is simply an infinite
sequence of unitary transformations, $U=\left\{  U_{n}\right\}  _{n\geq1}$.
\ We assume that each $U_{n}$\ acts on $p\left(  n\right)  $ qubits for some
known polynomial $p$. \ We also assume that given an $n$-bit string\ as input,
a quantum algorithm calls only $U_{n}$, not $U_{m}$\ for any $m\neq n$. \ This
assumption is only made for simplicity; our results would go through without
it.\footnote{If one made the analogous assumption in \textit{classical}
complexity---that given an input of length $n$, an algorithm can query the
oracle only on strings of length $n$---one could simplify a great many oracle
results without any loss of conceptual content.} \ When there is no danger of
confusion, we will refer to $U_{n}$\ simply as $U$.

Formally, the oracle access mechanism is as follows. \ Assume a quantum
computer's state has the form%
\[
\left\vert \Phi\right\rangle =\sum_{z}\alpha_{z}\left\vert z\right\rangle
\left\vert \phi_{z}\right\rangle ,
\]
where $\left\vert z\right\rangle $\ is a workspace register and $\left\vert
\phi_{b,z}\right\rangle $\ is a $p\left(  n\right)  $-qubit answer
register.\ \ Then to \textquotedblleft query $U_{n}$\textquotedblright\ means
to apply the $p\left(  n\right)  $-qubit unitary transformation that maps
$\left\vert \Phi\right\rangle $\ to%
\[
\left\vert \Phi^{\prime}\right\rangle =\sum_{z}\alpha_{z}\left\vert
z\right\rangle U_{n}\left\vert \phi_{z}\right\rangle .
\]
Let $\mathcal{C}$ be a quantum complexity class, and let $U=\left\{
U_{n}\right\}  _{n\geq1}$ be a quantum oracle. \ Then by $\mathcal{C}^{U}$, we
will mean the class of problems solvable by a $\mathcal{C}$ machine that,
given an input of length $n$, can query $U_{n}$\ at unit cost as many times as
it likes.

In defining the notion of quantum oracle, at least two choices present
themselves that have no counterpart for classical oracles:

\begin{enumerate}
\item[(1)] If we can apply a quantum oracle $U$,\ then can we also apply
controlled-$U$ (that is, $U$\ conditioned on a control qubit $\left\vert
b\right\rangle $)?

\item[(2)] If we can apply $U$, then can we also apply $U^{-1}$?
\end{enumerate}

At least for the present paper, the answers to these questions will not
matter, for the following reasons. \ First, all of the quantum oracles $U$
that we consider will be self-inverse (that is, $U=U^{-1}$). \ Second, while
our algorithms \textit{will} need to apply controlled-$U$, that is only for
the technical reason that we will define $U$\ so that $U\left\vert
\psi\right\rangle =-\left\vert \psi\right\rangle $\ if $\left\vert
\psi\right\rangle $\ is the marked state, and $U\left\vert \varphi
\right\rangle =\left\vert \varphi\right\rangle $\ whenever $\left\langle
\varphi|\psi\right\rangle =0$. \ If we stipulated instead that $U\left\vert
\psi\right\rangle \left\vert b\right\rangle =\left\vert \psi\right\rangle
\left\vert b\oplus1\right\rangle $\ and $U\left\vert \varphi\right\rangle
\left\vert b\right\rangle =\left\vert \varphi\right\rangle \left\vert
b\right\rangle $ whenever $\left\langle \varphi|\psi\right\rangle =0$, then
$U$\ alone would suffice.

Yet even though these choices will not matter for our results, it still seems
worthwhile to discuss them a bit, since they might arise in future work
involving quantum oracles.

One could argue that (i) the purpose of an oracle is to model a
\textit{subroutine} that an algorithm can call without understanding its
internal structure, and that (ii)\ given a quantum circuit for applying some
unitary operation $U$, one can easily produce a circuit for applying
controlled-$U$ or $U^{-1}$, without understanding anything about the original
circuit's structure. \ In particular, to produce a circuit for controlled-$U$,
one simply conditions each gate on the control qubit; while to produce a
circuit for $U^{-1}$, one simply inverts all the gates and reverses their
order. \ These considerations suggest that the answers to questions (1) and
(2) should both be `yes.' \ On the other hand, it would still be interesting
to know whether disallowing controlled-$U$ or $U^{-1}$\ would let us prove
more quantum oracle separations.\ \ (Note that if we disallow these
operations, then the set of inequivalent quantum oracles becomes
\textit{larger}.)

\section{Quantum Oracle Separations\label{QORACLE}}

The aim of this section is to prove Theorem \ref{qosepthm}: that there exists
a quantum oracle $U$ such that $\mathsf{QMA}^{U}\neq\mathsf{QCMA}^{U}$. \ The
same ideas will also yield a quantum oracle $V$ such that $\mathsf{BQP}^{V}
\mathsf{/qpoly}\neq\mathsf{BQP}^{V}\mathsf{/poly}$.

To prove these oracle separations, we first need a geometric lemma about
probability measures on quantum states. \ Let $\mu$\ be the uniform probability
measure over $N$-dimensional pure states\ (that is, over $\mathbb{CP}^{N-1}$).
\ The following notion will play a key role in our argument.

\begin{definition}
\label{puniform}For all $p\in\left[  0,1\right]  $, a probability measure
$\sigma$\ over $\mathbb{CP}^{N-1}$\ is called $p$\textit{-uniform}\ if
$p\sigma\leq\mu$. \ Equivalently, $\sigma$\ is $p$-uniform if it can be
obtained by starting from $\mu$,\ and then conditioning on an event that
occurs with probability at least $p$.
\end{definition}

So for example, we obtain a $p$-uniform measure if we start from $\mu$\ and
then condition on $\log_{2}1/p$\ bits of classical information about
$\left\vert \psi\right\rangle $. \ Our geometric lemma says that if
$\left\vert \psi\right\rangle $\ is drawn from a $p$-uniform measure, then for
every mixed state $\rho$, the squared fidelity between $\left\vert
\psi\right\rangle $\ and $\rho$\ has small expectation. \ More precisely:

\begin{lemma}
\label{sumip}Let $\sigma$\ be a $p$-uniform probability measure over
$\mathbb{CP}^{N-1}$. \ Then for all $\rho$,%
\[
\operatorname*{E}_{\left\vert \psi\right\rangle \in\sigma}\left[  \left\langle
\psi|\rho|\psi\right\rangle \right]  =O\left(  \frac{1+\log1/p}{N}\right)  .
\]

\end{lemma}

The proof of Lemma \ref{sumip} is deferred to Section \ref{LEMMA}. \ In this
section we assume the lemma, and show how to use it to prove our main result.
\ In particular, we show that any quantum algorithm needs $\Omega\left(
\sqrt{\frac{2^{n}}{m+1}}\right)  $\ queries to find an $n$-qubit marked state
$\left\vert \psi\right\rangle $, even if given $m$ bits of classical advice
about $\left\vert \psi\right\rangle $.

\begin{theorem}
\label{qosep}Suppose we are given oracle access to an $n$-qubit unitary $U$,
and want to decide which of the following holds:

\begin{itemize}
\item[(i)] There exists an $n$-qubit \textquotedblleft quantum marked
state\textquotedblright\ $\left\vert \psi\right\rangle $\ such that
$U\left\vert \psi\right\rangle =-\left\vert \psi\right\rangle $, but
$U\left\vert \phi\right\rangle =\left\vert \phi\right\rangle $\ whenever
$\left\langle \phi|\psi\right\rangle =0$; or

\item[(ii)] $U=I$ is the identity operator.
\end{itemize}

Then even if we have an $m$-bit classical witness $w$\ in support of case (i),
we still need $\Omega\left(  \sqrt{\frac{2^{n}}{m+1}}\right)  $\ queries to
verify the witness, with bounded probability of error.
\end{theorem}

\begin{proof}
If $m=\Omega\left(  2^{n}\right)  $\ then the theorem is certainly true, so
suppose $m=o\left(  2^{n}\right)  $. \ Let $A$ be a quantum algorithm that
queries $U$. \ Also, let $U_{\psi}$\ be an $n$-qubit unitary such that
$U_{\psi}\left\vert \psi\right\rangle =-\left\vert \psi\right\rangle $, but
$U_{\psi}\left\vert \phi\right\rangle =\left\vert \phi\right\rangle
$\ whenever $\left\langle \phi|\psi\right\rangle =0$. \ Then $A$'s goal is to
accept if and only if $U=U_{\psi}$ for some $\left\vert \psi\right\rangle $.

For each $n$-qubit pure state $\left\vert \psi\right\rangle $, let us fix a
classical witness $w\in\left\{  0,1\right\}  ^{m}$\ that maximizes the
probability that $A$ accepts, given $U_{\psi}$\ as oracle. Let $S\left(
w\right)  $\ be the set of $\left\vert \psi\right\rangle $'s associated with a
given witness $w$. \ Since the $S\left(  w\right)  $'s form a partition of
$\mathbb{CP}^{2^{n}-1}$, clearly there exists a witness, call it $w^{\ast}$,
such that%
\[
\Pr_{\left\vert \psi\right\rangle \in\mu}\left[  \left\vert \psi\right\rangle
\in S\left(  w^{\ast}\right)  \right]  \geq\frac{1}{2^{m}}.
\]
Fix that $w^{\ast}$ (or in other words, hardwire $w^{\ast}$\ into $A$). \ Then
to prove the theorem, it suffices to establish the following claim: $A$ cannot
distinguish the case $U=U_{\psi}$\ from the case $U=I$\ by making $o\left(
\sqrt{\frac{2^{n}}{m+1}}\right)  $\ queries to $U$, with high probability if
$\left\vert \psi\right\rangle $\ is chosen uniformly at random from $S\left(
w^{\ast}\right)  $.

To prove the claim, we use a generalization of the hybrid argument of Bennett
et al. \cite{bbbv}. \ Suppose that $A$ makes $T$ queries to $U$.
\ (Technically speaking, we should also allow queries to controlled-$U$, but
this will make no difference in our analysis.) \ Then for all $0\leq t\leq T$,
let $\left\vert \Phi_{t}\right\rangle $\ be the final state of $A$, assuming
that $U=I$\ for the first $t$ queries, and $U=U_{\psi}$\ for the remaining
$T-t$\ queries. \ Thus $\left\vert \Phi_{0}\right\rangle $\ is the final state
in case (i), while $\left\vert \Phi_{T}\right\rangle $\ is the final state in
case (ii). \ We will argue that $\left\vert \Phi_{t}\right\rangle $\ cannot be
very far from $\left\vert \Phi_{t-1}\right\rangle $, with high probability
over the choice of marked state $\left\vert \psi\right\rangle $.
\ Intuitively, this is because the computations of $\left\vert \Phi
_{t}\right\rangle $\ and $\left\vert \Phi_{t-1}\right\rangle $ differ in only
a single query, and with high probability that query cannot have much overlap
with $\left\vert \psi\right\rangle $. \ We will then conclude, by the triangle
inequality, that $\left\vert \Phi_{0}\right\rangle $\ cannot be far from
$\left\vert \Phi_{T}\right\rangle $\ unless $T$ is large.

More formally, let $\rho_{t}$\ be the marginal state of the query register
just before the $t^{th}$\ query, assuming the \textquotedblleft control
case\textquotedblright\ $U=I$. \ Also, let $\rho_{t}=\sum p_{i}\left\vert
\varphi_{i}\right\rangle \left\langle \varphi_{i}\right\vert $\ be an
arbitrary decomposition of $\rho_{t}$ into pure states. \ Then for every $i$,
the component of $\left\vert \varphi_{i}\right\rangle $ orthogonal to
$\left\vert \psi\right\rangle $\ is unaffected by the $t^{th}$\ query.
\ Therefore%
\[ \left\Vert \left\vert \Phi_{t}\right\rangle -\left\vert \Phi_{t-1}%
\right\rangle \right\Vert _{2}  \leq\sum_{i}p_{i}\cdot2\left\vert
\left\langle \varphi_{i}|\psi\right\rangle \right\vert
  =2\sum_{i}p_{i}\sqrt{\left\langle \psi|\varphi_{i}\right\rangle
\left\langle \varphi_{i}|\psi\right\rangle }
  \leq2\sqrt{\sum_{i}p_{i}\left\langle \psi|\varphi_{i}\right\rangle
\left\langle \varphi_{i}|\psi\right\rangle }
  =2\sqrt{\left\langle \psi|\rho_{t}|\psi\right\rangle },
\]
where the third line uses the Cauchy-Schwarz inequality (the average of the
square root is at most the square root of the average). \ Now let $\sigma$\ be
the uniform probability measure over $S\left(  w^{\ast}\right)  $, and observe
that $\sigma$\ is $2^{-m}$-uniform. \ So by Lemma \ref{sumip},%
\[
\operatorname*{E}_{\left\vert \psi\right\rangle \in\sigma}\left[  \left\Vert
\left\vert \Phi_{t}\right\rangle -\left\vert \Phi_{t-1}\right\rangle
\right\Vert _{2}\right] \leq2\operatorname*{E}_{\left\vert \psi
\right\rangle \in\sigma}\left[  \sqrt{\left\langle \psi|\rho_{t}%
|\psi\right\rangle }\right]
  \leq2\sqrt{\operatorname*{E}_{\left\vert \psi\right\rangle \in\sigma
}\left[  \left\langle \psi|\rho_{t}|\psi\right\rangle \right]  }
  \leq2\sqrt{\frac{1+\ln\left(  1/2^{-m}\right)  }{2^{n}}}
  =O\left(  \sqrt{\frac{m+1}{2^{n}}}\right)  ,
\]
where the second line again uses the Cauchy-Schwarz inequality. \ Finally,%
\[
\operatorname*{E}_{\left\vert \psi\right\rangle \in S\left(  w^{\ast}\right)
}\left[  \left\Vert \left\vert \Phi_{T}\right\rangle -\left\vert \Phi
_{0}\right\rangle \right\Vert _{2}\right]  \leq\sum_{t=1}^{T}\operatorname*{E}%
_{\left\vert \psi\right\rangle \in S\left(  w^{\ast}\right)  }\left[
\left\Vert \left\vert \Phi_{t}\right\rangle -\left\vert \Phi_{t-1}%
\right\rangle \right\Vert _{2}\right]  =O\left(  T\sqrt{\frac{m+1}{2^{n}}%
}\right)
\]
by the triangle inequality. \ This implies that, for $\left\vert \Phi
_{T}\right\rangle $\ and $\left\vert \Phi_{0}\right\rangle $\ to be
distinguishable with $\Omega\left(  1\right)  $ bias, we must have
$T=\Omega\left(  \sqrt{\frac{2^{n}}{m+1}}\right)  $.
\end{proof}

Using Theorem \ref{qosep}, we can straightforwardly show a quantum oracle
separation between $\mathsf{QMA}$\ and $\mathsf{QCMA}$.

\begin{proof}
[Proof of Theorem \ref{qosepthm}]Let $L$ be a unary language chosen uniformly
at random. \ The oracle $U=\left\{  U_{n}\right\}  _{n\geq1}$ is as follows:
if $0^{n}\in L$, then $U_{n}\left\vert \psi_{n}\right\rangle =-\left\vert
\psi_{n}\right\rangle $\ for some $n$-qubit marked state $\left\vert \psi
_{n}\right\rangle $ chosen uniformly at random, while $U_{n}\left\vert
\varphi\right\rangle =\left\vert \varphi\right\rangle $\ whenever
$\left\langle \varphi|\psi_{n}\right\rangle =0$. \ Otherwise, if $0^{n}\notin
L$, then $U_{n}$\ is the $n$-qubit identity operation.

Almost by definition, $L\in\mathsf{QMA}^{U}$.\ \ For given a quantum witness
$\left\vert \varphi\right\rangle $, the $\mathsf{QMA}$\ verifier first\ prepares
the state $\frac{1}{\sqrt{2}}\left(  \left\vert 0\right\rangle \left\vert
\varphi\right\rangle +\left\vert 1\right\rangle \left\vert \varphi\right\rangle
\right)  $, then applies $U_{n}$\ to the second register conditioned on the
first register being $\left\vert 1\right\rangle $. \ Next the verifier applies a
Hadamard gate to the first register, measures it, and accepts if and only if
$\left\vert 1\right\rangle $\ is observed. \ If $0^{n}\in L$, then there exists
a witness---namely $\left\vert \varphi \right\rangle =\left\vert
\psi_{n}\right\rangle $---that causes the verifier to accept with probability
$1$. \ On the other hand, if $0^{n}\notin L$, then\ \textit{no} witness causes
the verifier to accept with nonzero probability.

On the other hand, we claim that $L\notin\mathsf{QCMA}^{U}$\ with probability
$1$ over the choice of $L$ and $U$.\ \ This can be seen as follows.\ \ Fix a
$\mathsf{QCMA}$ machine $M$, and let $S_{M}\left(  n\right)  $\ be the event
that $M^{U}$ \textit{succeeds} on $0^{n}$: that is, either $0^{n}\in L$\ and
there exists a string $w$ such that $M^{U}$ accepts $\left\vert 0^{n}%
\right\rangle \left\vert w\right\rangle $\ with probability at least $2/3$, or
$0^{n}\notin L$\ and $M^{U}$ accepts $\left\vert 0^{n}\right\rangle \left\vert
w\right\rangle $\ with probability at most $1/3$ for all $w$. \ Then Theorem
\ref{qosep}\ readily implies that there exists a positive integer $N$\ such
that for all $n\geq N$,%
\[
\Pr_{L,U}\left[  S_{M}\left(  n\right)  ~|~S_{M}\left(  1\right)
,\ldots,S_{M}\left(  n-1\right)  \right]  \leq\frac{2}{3}.
\]
Hence%
\[
\Pr_{L,U}\left[  S_{M}\left(  1\right)  \wedge S_{M}\left(  2\right)
\wedge\cdots\right]  =0.
\]
Now, because of the Solovay-Kitaev Theorem \cite{kitaev:ec}, the number of
possible $\mathsf{QCMA}$\ machines is only countably infinite. \ So by the
union bound,%
\[
\Pr_{L,U}\left[  \exists M:S_{M}\left(  1\right)  \wedge S_{M}\left(
2\right)  \wedge\cdots\right]  =0
\]
as well.
\end{proof}

We can similarly show a quantum oracle separation between $\mathsf{BQP/qpoly}%
$\ and $\mathsf{BQP/poly}$.

\begin{theorem}
\label{qpoly}There exists a quantum oracle $U$ such that $\mathsf{BQP}%
^{U}\mathsf{/qpoly}\neq\mathsf{BQP}^{U}\mathsf{/poly}$.
\end{theorem}

\begin{proof}
In this case $U_{n}$ will act on $2n$\ qubits. \ Let $L$ be a binary language
chosen uniformly at random, and let $L\left(  x\right)  =1$\ if $x\in L$\ and
$L\left(  x\right)  =0$ otherwise. \ Also, for all $n$, let $\left\vert
\psi_{n}\right\rangle $\ be an $n$-qubit state chosen uniformly at random.
\ Then $U_{n}$\ acts as follows: for all $x\in\left\{  0,1\right\}  ^{n}$,%
\[
U_{n}\left(  \left\vert \psi_{n}\right\rangle \left\vert x\right\rangle
\right)  =\left(  -1\right)  ^{L\left(  x\right)  }\left\vert \psi
_{n}\right\rangle \left\vert x\right\rangle ,
\]
but $U_{n}\left(  \left\vert \phi\right\rangle \left\vert x\right\rangle
\right)  =\left\vert \phi\right\rangle \left\vert x\right\rangle $\ whenever
$\left\langle \phi|\psi_{n}\right\rangle =0$. \ Clearly $L\in\mathsf{BQP}%
^{U}\mathsf{/qpoly}$; we just take $\left\vert \psi_{n}\right\rangle $ as the
advice. \ On the other hand, by essentially the same argument as for Theorem
\ref{qosepthm}, one can show that $L\notin\mathsf{BQP}^{U}\mathsf{/poly}%
$\ with probability $1$ over $L$ and $U$.
\end{proof}

\subsection{Proof of Geometric Lemma\label{LEMMA}}

In this section we fill in the proof of Lemma \ref{sumip}, thereby completing
the oracle separation theorems.

In proving Lemma \ref{sumip}, the first step is to ask the following question:
among all $p$-uniform probability measures $\sigma$, which is the one that
maximizes\ $\operatorname*{E}_{\left\vert \psi\right\rangle \in\sigma}\left[
\left\vert \left\langle \psi|0\right\rangle \right\vert ^{2}\right]  $? \ We
can think of the set of quantum states $\mathbb{CP}^{N-1}$\ as a container,
which contains a fluid $\sigma$\ that is gravitationally attracted to the
state $\left\vert 0\right\rangle $. \ Then intuitively, the answer is clear:
the way to maximize $\operatorname*{E}_{\left\vert \psi\right\rangle \in
\sigma}\left[  \left\vert \left\langle \psi|0\right\rangle \right\vert
^{2}\right]  $ is to\ \textquotedblleft fill the container from the
bottom,\textquotedblright\ subject to the density constraint $p\sigma\leq\mu$.
\ In other words, the optimal $\sigma$\ should be the uniform measure over the
region $\mathcal{R}\left(  p\right)  $\ given by $\left\vert \left\langle
\psi|0\right\rangle \right\vert \geq h\left(  p\right)  $, where $h\left(
p\right)  $\ is chosen so that the volume of $\mathcal{R}\left(  p\right)
$\ is a $p$ fraction of the total volume of $\mathbb{CP}^{N-1}$. \ The
following lemma makes this intuition rigorous.

\begin{lemma}
\label{gravity}Among all $p$-uniform probability measures $\sigma$\ over
$\mathbb{CP}^{N-1}$, the one that maximizes $\operatorname*{E}_{\left\vert
\psi\right\rangle \in\sigma}\left[  \left\vert \left\langle \psi
|0\right\rangle \right\vert ^{2}\right]  $\ is $\tau\left(  p\right)  $, the
uniform measure over the region $\mathcal{R}\left(  p\right)  $ defined above.
\end{lemma}

\begin{proof}
Since $\left\vert \left\langle \psi|0\right\rangle \right\vert ^{2}$\ is
nonnegative, we can write%
\[
\operatorname*{E}_{\left\vert \psi\right\rangle \in\sigma}\left[  \left\vert
\left\langle \psi|0\right\rangle \right\vert ^{2}\right]  =\int_{0}^{\infty
}\Pr_{\left\vert \psi\right\rangle \in\sigma}\left[  \left\vert \left\langle
\psi|0\right\rangle \right\vert ^{2}\geq y\right]  dy.
\]
We claim that setting $\sigma:=\tau\left(  p\right)  $\ maximizes the
integrand for every value of $y$. \ Certainly, then, setting $\sigma
:=\tau\left(  p\right)  $\ maximizes the integral itself as well.

To prove the claim, we consider two cases. \ First, if $y\leq h\left(
p\right)  ^{2}$, then%
\[
\Pr_{\left\vert \psi\right\rangle \in\tau\left(  p\right)  }\left[  \left\vert
\left\langle \psi|0\right\rangle \right\vert ^{2}\geq y\right]  =1,
\]
which is certainly maximal. \ Second, if $y>h\left(  p\right)  ^{2}$, then%
\[
\Pr_{\left\vert \psi\right\rangle \in\tau\left(  p\right)  }\left[  \left\vert
\left\langle \psi|0\right\rangle \right\vert ^{2}\geq y\right]  =\frac{1}%
{p}\cdot\Pr_{\left\vert \psi\right\rangle \in\mu}\left[  \left\vert
\left\langle \psi|0\right\rangle \right\vert ^{2}\geq y\right]  .
\]
This is maximal as well, since%
\[
\Pr_{\left\vert \psi\right\rangle \in\sigma}\left[  \left\vert \left\langle
\psi|0\right\rangle \right\vert ^{2}\geq y\right]  \leq\frac{1}{p}\cdot
\Pr_{\left\vert \psi\right\rangle \in\mu}\left[  \left\vert \left\langle
\psi|0\right\rangle \right\vert ^{2}\geq y\right]  .
\]
for all $p$-uniform probability measures $\sigma$.
\end{proof}

Lemma \ref{gravity}\ completely describes the probability measure that
maximizes $\operatorname*{E}_{\left\vert \psi\right\rangle \in\sigma}\left[
\left\vert \left\langle \psi|0\right\rangle \right\vert ^{2}\right]  $, except
for one detail: the value of $h\left(  p\right)  $ (or equivalently, the
radius of $\mathcal{R}\left(  p\right)  $). \ The next lemma completes the
picture.

\begin{lemma}
\label{hball}For all $p$,%
\[
h\left(  p\right)  =\sqrt{1-p^{1/\left(  N-1\right)  }}=\Theta\left(
\sqrt{\frac{\log1/p}{N}}\right)  .
\]

\end{lemma}

\begin{proof}
We will show that for all $h$,%
\[
\Pr_{\left\vert \psi\right\rangle \in\mu}\left[  \left\vert \left\langle
\psi|0\right\rangle \right\vert \geq h\right]  =\left(  1-h^{2}\right)
^{N-1},
\]
where $\mu$\ is the uniform probability measure over $\mathbb{CP}^{N-1}$.
\ Setting $p:=\Pr_{\left\vert \psi\right\rangle \in\mu}\left[  \left\vert
\left\langle \psi|0\right\rangle \right\vert \geq h\right]  $\ and solving for
$h$ then yields the lemma.

Let $\overrightarrow{z}=\left(  z_{0},\ldots,z_{N-1}\right)  $\ be a complex
vector; then let $\overrightarrow{r}=\left(  r_{0},\ldots,r_{N-1}\right)
$\ and $\overrightarrow{\theta}=\left(  \theta_{0},\ldots,\theta_{N-1}\right)
$\ be real vectors such that $z_{k}=r_{k}e^{i\theta_{k}}$ for each coordinate
$k$. \ Also, let $\mathcal{D}$ be a Gaussian probability measure on
$\mathbb{C}^{N}$, with density function%
\[
P\left(  \overrightarrow{z}\right)  =P\left(  \overrightarrow{r}\right)
=\frac{1}{\pi^{N}}e^{-\left\Vert \overrightarrow{r}\right\Vert _{2}^{2}}.
\]
Let $d\overrightarrow{r}$\ be shorthand for $dr_{0}\cdots dr_{N-1}$. \ Then we
can express the probability that $\left\vert \left\langle \psi|0\right\rangle
\right\vert \geq h$ as%
\begin{align*}
\Pr_{\left\vert \psi\right\rangle \in\mu}\left[  \left\vert \left\langle
\psi|0\right\rangle \right\vert \geq h\right]   &  =\Pr_{\overrightarrow{z}%
\in\mathcal{D}}\left[  \left\vert z_{0}\right\vert \geq h\left\Vert
\overrightarrow{z}\right\Vert _{2}\right] \\
&  =\Pr_{\overrightarrow{r},\overrightarrow{\theta}}\left[  r_{0}\geq
h\left\Vert \overrightarrow{r}\right\Vert _{2}\right] \\
&  =\int_{\overrightarrow{r},\overrightarrow{\theta}~:~r_{0}\geq h\left\Vert
\overrightarrow{r}\right\Vert _{2}}P\left(  \overrightarrow{r}\right)
~r_{0}\cdots r_{N-1}~d\overrightarrow{r}d\overrightarrow{\theta}\\
&  =\left(  2\pi\right)  ^{N}\int_{\overrightarrow{r}~:~r_{0}\geq h\left\Vert
\overrightarrow{r}\right\Vert _{2}}\frac{1}{\pi^{N}}e^{-\left\Vert
\overrightarrow{r}\right\Vert _{2}^{2}}~r_{0}\cdots r_{N-1}~d\overrightarrow
{r}\\
&  =\int_{r_{1},\ldots,r_{N-1}=0}^{\infty}\left(  \int_{r_{0}=h\sqrt
{\frac{r_{1}^{2}+\cdots+r_{N-1}^{2}}{1-h^{2}}}}^{\infty}2e^{-r_{0}^{2}}%
r_{0}dr_{0}\right)  2^{N-1}e^{-r_{1}^{2}-\cdots-r_{N-1}^{2}}~r_{1}dr_{1}\cdots
r_{N-1}dr_{N-1}\\
&  =\int_{r_{1},\ldots,r_{N-1}=0}^{\infty}e^{-\left(  r_{1}^{2}+\cdots
+r_{N-1}^{2}\right)  \cdot h^{2}/\left(  1-h^{2}\right)  }2^{N-1}e^{-r_{1}%
^{2}-\cdots-r_{N-1}^{2}}~r_{1}dr_{1}\cdots r_{N-1}dr_{N-1}\\
&  =\int_{r_{1},\ldots,r_{N-1}=0}^{\infty}2^{N-1}e^{-\left(  r_{1}^{2}%
+\cdots+r_{N-1}^{2}\right)  /\left(  1-h^{2}\right)  }~r_{1}dr_{1}\cdots
r_{N-1}dr_{N-1}\\
&  =\left(  \int_{r=0}^{\infty}2e^{-r^{2}/\left(  1-h^{2}\right)  }rdr\right)
^{N-1}\\
&  =\left(  1-h^{2}\right)  ^{N-1}.
\end{align*}

\end{proof}

By combining Lemmas \ref{gravity} and \ref{hball}, we can now prove Lemma
\ref{sumip}: that if $\sigma$\ is $p$-uniform, then for all mixed states
$\rho$,%
\[
\operatorname*{E}_{\left\vert \psi\right\rangle \in\sigma}\left[  \left\langle
\psi|\rho|\psi\right\rangle \right]  =O\left(  \frac{1+\log1/p}{N}\right)  .
\]

\begin{proof}
[Proof of Lemma \ref{sumip}]If $p\leq e^{-\Omega\left(  N\right)  }$\ then the
lemma is certainly true, so suppose $p\geq e^{-O\left(  N\right)  }$. \ Since
the concluding inequality is linear in $\rho$, we can assume without loss of
generality that $\rho$\ is a pure state. \ Indeed, by symmetry we can assume
that $\rho=\left\vert 0\right\rangle \left\langle 0\right\vert $. \ So our aim
is to upper-bound $\operatorname*{E}_{\left\vert \psi\right\rangle \in\sigma
}\left[  \left\vert \left\langle \psi|0\right\rangle \right\vert ^{2}\right]
$, where $\sigma$\ is any $p$-uniform probability measure.\ \ By Lemma
\ref{gravity}, we can assume without loss of generality that $\sigma
=\tau\left(  p\right)  $ is the uniform measure over all $\left\vert
\psi\right\rangle $\ such that $\left\vert \left\langle \psi|0\right\rangle
\right\vert \geq h\left(  p\right)  $. \ Then letting%
\begin{align*}
\left\vert \psi\right\rangle  &  =\alpha_{0}\left\vert 0\right\rangle
+\cdots+\alpha_{N-1}\left\vert N-1\right\rangle ,\\
r  &  =\sqrt{\left\vert \alpha_{1}\right\vert ^{2}+\cdots+\left\vert
\alpha_{N-1}\right\vert ^{2}},
\end{align*}
we have%
\begin{align*}
\operatorname*{E}_{\left\vert \psi\right\rangle \in\tau\left(  p\right)
}\left[  \left\vert \left\langle \psi|0\right\rangle \right\vert ^{2}\right]
&  =\operatorname*{E}_{\left\vert \psi\right\rangle ~:~\left\vert \alpha
_{0}\right\vert \geq h\left(  p\right)  }\left[  \left\vert \alpha
_{0}\right\vert ^{2}\right] \\
&  =\operatorname*{E}_{\left\vert \psi\right\rangle ~:~r^{2}\leq1-h\left(
p\right)  ^{2}}\left[  1-r^{2}\right] \\
&  =\frac{\int_{0}^{\sqrt{1-h\left(  p\right)  ^{2}}}r^{2N-3}\left(
1-r^{2}\right)  dr}{\int_{0}^{\sqrt{1-h\left(  p\right)  ^{2}}}r^{2N-3}dr}\\
&  =\frac{\left[  \frac{r^{2N-2}}{2N-2}-\frac{r^{2N}}{2N}\right]  _{0}%
^{\sqrt{1-h\left(  p\right)  ^{2}}}}{\left[  \frac{r^{2N}}{2N}\right]
_{0}^{\sqrt{1-h\left(  p\right)  ^{2}}}}\\
&  =\frac{1-\left(  1-\frac{1}{N}\right)  \left(  1-h\left(  p\right)
^{2}\right)  }{\left(  1-\frac{1}{N}\right)  \left(  1-h\left(  p\right)
^{2}\right)  }\\
&  =O\left(  \frac{1}{N}+h\left(  p\right)  ^{2}\right) \\
&  =O\left(  \frac{1+\log1/p}{N}\right)  ,
\end{align*}
where the last line follows from Lemma \ref{hball}.
\end{proof}

\section{Upper Bound\label{UPPER}}

In this section we show that the lower bound of Theorem \ref{qosep}\ is
basically tight. \ In particular, let $U$ be an $n$-qubit quantum oracle,\ and
suppose we are given an $m$-bit classical proof that $U$ is not the identity,
but instead conceals a marked state $\left\vert \psi\right\rangle $ such that
$U\left\vert \psi\right\rangle =-\left\vert \psi\right\rangle $. \ Then
provided $2n\leq m\leq2^{n}$, a quantum algorithm can verify the proof by
making $O\left(  \sqrt{2^{n}/m}\right)  $\ oracle calls to $U$. \ This matches
our lower bound when $m\geq2n$.\footnote{When $m\ll2n$, the best upper bound
we know is the trivial $O\left(  \sqrt{2^{n}}\right)  $. \ However, we
conjecture that $O\left(  \sqrt{2^{n}/m}\right)  $\ is achievable in this case
as well.}

Let $N=2^{n}$\ be the dimension of $U$'s Hilbert space. \ Then the idea of our
algorithm is to use a \textquotedblleft mesh\textquotedblright\ of states
$\left\vert \phi_{1}\right\rangle ,\ldots,\left\vert \phi_{M}\right\rangle
\in\mathbb{CP}^{N-1}$, at least one of which has nontrivial overlap with every
pure state in $\mathbb{CP}^{N-1}$. \ A classical proof can then help the
algorithm by telling it the $\left\vert \phi_{i}\right\rangle $\ that is
closest to $\left\vert \psi\right\rangle $. \ More formally, define the
$h$\textit{-ball} about $\left\vert \phi\right\rangle $\ to be the set of
$\left\vert \varphi\right\rangle $\ such that $\left\vert \left\langle
\phi|\varphi\right\rangle \right\vert \geq h$. \ Then define an $h$%
\textit{-net for }$\mathbb{CP}^{N-1}$\textit{ of size }$M$\ to be a set of
states $\left\vert \phi_{1}\right\rangle ,\ldots,\left\vert \phi
_{M}\right\rangle $ such that every $\left\vert \psi\right\rangle
\in\mathbb{CP}^{N-1}$ is contained in the $h$-ball about $\left\vert \phi
_{i}\right\rangle $\ for some $i$.\footnote{These objects are often called
$\varepsilon$-nets, with the obvious relation $h=\cos\varepsilon$.} \ We will
use the following theorem, which follows from Corollary 1.2 of
B\"{o}r\"{o}czky and Wintsche\ \cite{boroczky}.

\begin{theorem}
[\cite{boroczky}]\label{hnet}For all $0<h<1$, there exists an $h$-net for
$\mathbb{CP}^{N-1}$\ of size%
\[
O\left(  \frac{N^{3/2}\log\left(  2+Nh^{2}\right)  }{\left(  1-h^{2}\right)
^{N}}\right)  .
\]

\end{theorem}

B\"{o}r\"{o}czky and Wintsche\ do not provide an explicit construction of such
an $h$-net;\ they only prove that it exists.\footnote{Note that we cannot just
start from an explicit construction of a sphere-packing, and then double the
radius of the spheres to get a covering. \ We could do that if we wanted a
covering of $\mathbb{CP}^{N-1}$\ by \textit{small} balls. \ But in our case,
$h$ is close to zero, which means that the balls already have close to the
maximal radius.} \ Later, we will give an explicit construction with only
slightly worse parameters than those of Theorem \ref{hnet}. \ But first, let
us prove an upper bound on query complexity.

\begin{theorem}
\label{upperbound}Suppose we have an $n$-qubit quantum oracle $U$ such that
either (i) $U=U_{\psi}$\ for some $\left\vert \psi\right\rangle $, or (ii)
$U=I$\ is the identity operator. \ Then given an $m$-bit classical witness in
support of case (i), where $m\geq2n$, there exists a quantum algorithm that
verifies the witness using $O\left(  \sqrt{2^{n}/m}+1\right)  $\ queries to
$U$.
\end{theorem}

\begin{proof}
By Theorem \ref{hnet},\ there exists an $h$-net $\mathcal{S}$\ for
$\mathbb{CP}^{2^{n}-1}$\ of cardinality%
\[
\left\vert \mathcal{S}\right\vert =O\left(  \frac{2^{3n/2}\log\left(
2+2^{n}h^{2}\right)  }{\left(  1-h^{2}\right)  ^{2^{n}}}\right)  .
\]
Setting $\left\vert \mathcal{S}\right\vert =2^{m}$\ gives%
\[
m\leq\frac{3n}{2}+2^{n}\log\left(  \frac{1}{1-h^{2}}\right)  +O\left(  \log
n\right)  .
\]
Solving for $h$, we obtain%
\[
h\geq\sqrt{\frac{m-3n/2-O\left(  \log n\right)  }{2^{n}}},
\]
which is $\Omega\left(  \sqrt{m/2^{n}}\right)  $\ provided $m\geq2n$. \ So
there exists a collection of $M=2^{m}$\ states, $\left\vert \phi
_{1}\right\rangle ,\ldots,\left\vert \phi_{M}\right\rangle \in\mathbb{CP}%
^{2^{n}-1}$, such that for every $\left\vert \psi\right\rangle $, there exists
an $i$\ such that $\left\vert \left\langle \phi_{i}|\psi\right\rangle
\right\vert \geq h$ where $h=\Omega\left(  \sqrt{m/2^{n}}\right)  $.

Given an oracle $U=U_{\left\vert \psi\right\rangle }$, the classical witness
$w\in\left\{  0,1\right\}  ^{m}$ will simply encode an index $i$ such that
$\left\vert \left\langle \phi_{i}|\psi\right\rangle \right\vert \geq h$. \ If
we prepare $\left\vert \phi_{i}\right\rangle $\ and feed it to $U$, then the
probability of finding the marked state $\left\vert \psi\right\rangle $ is
$\left\vert \left\langle \phi_{i}|\psi\right\rangle \right\vert ^{2}\geq
h^{2}$. \ Furthermore, if we do find $\left\vert \psi\right\rangle $, we will
know we did (i.e. a control qubit will be $\left\vert 1\right\rangle
$\ instead of $\left\vert 0\right\rangle $). \ From these facts, it follows
immediately from the amplitude amplification theorem of Grover
\cite{grover:framework}\ and Brassard et al. \cite{bhmt} that we can find
$\left\vert \psi\right\rangle $ with probability $\Omega\left(  1\right)
$\ using%
\[
O\left(  \sqrt{\frac{1}{h^{2}}}+1\right)  =O\left(  \sqrt{\frac{2^{n}}{m}%
}+1\right)
\]
queries to $U$.
\end{proof}

Of course, if we care about \textit{computational} complexity as well as query
complexity, then it is not enough for an $h$-net to exist---we also need the
states in the $h$-net to be efficiently preparable. \ Fortunately, proving an
explicit version of Theorem \ref{hnet}\ turns out to be simpler than one might
expect. \ We will do so with the help of the following inequality.

\begin{lemma}
\label{ineq}Let $x_{1}\geq\cdots\geq x_{N}\geq0$\ be nonnegative real numbers
with $x_{1}^{2}+\cdots+x_{N}^{2}=1$. \ Then for all $k\in\left\{
1,\ldots,N\right\}  $,%
\[
\max_{1\leq t\leq k}\left[  \frac{x_{1}+\cdots+x_{t}}{\sqrt{t}}\right]
\geq\sqrt{\frac{k}{N\left\lceil \log_{2}N\right\rceil }}.
\]

\end{lemma}

\begin{proof}
Let $L=\left\lceil \log_{2}N\right\rceil $. \ Then for all $i\in\left\{
1,\ldots,L\right\}  $, let $s_{i}=x_{2^{i-1}}^{2}+\cdots+x_{2^{i}-1}^{2}$,
where we adopt the convention that $x_{j}=0$\ if $j>N$. \ Then%
\[
s_{1}+\cdots+s_{L}=x_{1}^{2}+\cdots+x_{N}^{2}=1,
\]
so certainly there exists an $i\in\left\{  1,\ldots,L\right\}  $\ such that
$s_{i}\geq1/L$. \ Fix that $i$. \ Then since the $x_{j}$'s\ are arranged in
nonincreasing order, we have%
\[
x_{2^{i-1}}\geq\sqrt{\frac{s_{i}}{2^{i-1}}}\geq\sqrt{\frac{1}{2^{i-1}L}}.
\]
There are now two cases. \ First, if $k\leq2^{i-1}$ then%
\[
\max_{1\leq t\leq k}\left[  \frac{x_{1}+\cdots+x_{t}}{\sqrt{t}}\right]
\geq\frac{x_{1}+\cdots+x_{k}}{\sqrt{k}}\geq\frac{k}{\sqrt{k}}x_{2^{i-1}}%
\geq\sqrt{\frac{k}{2^{i-1}L}}\geq\sqrt{\frac{k}{N\left\lceil \log
_{2}N\right\rceil }}.
\]
Second, if $2^{i-1}\leq k$\ then%
\[
\max_{1\leq t\leq k}\left[  \frac{x_{1}+\cdots+x_{t}}{\sqrt{t}}\right]
\geq\frac{x_{1}+\cdots+x_{2^{i-1}}}{\sqrt{2^{i-1}}}\geq\frac{2^{i-1}}%
{\sqrt{2^{i-1}}}x_{2^{i-1}}\geq\sqrt{\frac{1}{L}}\geq\sqrt{\frac
{k}{N\left\lceil \log_{2}N\right\rceil }}.
\]
This completes the proof.\footnote{One might wonder whether the $\sqrt
{1/\left\lceil \log_{2}N\right\rceil }$\ factor can be eliminated. \ However,
a simple example shows that it can be improved by at most a constant factor.
\ Suppose $x_{j}:=\sqrt{\frac{1}{jw}}$, where $w=\sum_{j=1}^{n}\frac{1}%
{j}\approx\ln N$. \ Then for all $t\in\left\{  1,\ldots,N\right\}  $, we have%
\[
\frac{x_{1}+\cdots+x_{t}}{\sqrt{t}}\approx\frac{2}{\sqrt{\ln N}}.
\]
}
\end{proof}

We now use Lemma \ref{ineq}\ to construct an $h$-net.

\begin{theorem}
\label{explicit}For all $0<h<1$, there exists an $h$-net $\left\vert \phi
_{1}\right\rangle ,\ldots,\left\vert \phi_{M}\right\rangle $\ for
$\mathbb{CP}^{N-1}$\ of size $M=4N\cdot2^{O\left(  h^{2}N\log^{2}N\right)  }$,
as well as a quantum algorithm that runs in time polynomial in $\log M$\ and
that prepares the state $\left\vert \phi_{i}\right\rangle $\ given $i$ as input.
\end{theorem}

\begin{proof}
Assume without loss of generality that $N=2^{n}$\ and $M=2^{m}$ are both
powers of $2$, and\ let $\left\vert \psi\right\rangle $\ be an $n$-qubit
target state. \ Then it suffices to show that a quantum algorithm, using%
\[
m=\log_{2}M=n+2+O\left(  h^{2}2^{n}n^{2}\right)
\]
bits of classical advice, can prepare a state\ $\left\vert \phi\right\rangle
$\ such that $\left\vert \left\langle \phi|\psi\right\rangle \right\vert \geq
h$ in time polynomial in $m$.

Let $k:=\left\lfloor \frac{m}{n+2}\right\rfloor $. \ Also, let us express
$\left\vert \psi\right\rangle $\ in the computational basis as%
\[
\left\vert \psi\right\rangle =\sum_{z\in\left\{  1,\ldots,N\right\}  }%
\alpha_{z}\left\vert z\right\rangle ,
\]
and let $\left\vert z_{1}\right\rangle ,\ldots,\left\vert z_{N}\right\rangle
$\ be an ordering of basis states\ with the property that $\left\vert
\alpha_{z_{1}}\right\vert \geq\cdots\geq\left\vert \alpha_{z_{N}}\right\vert
$. \ Then by Lemma \ref{ineq},\ there exists an integer $t\in\left\{
1,\ldots,k\right\}  $\ such that%
\[
\frac{\left\vert \alpha_{z_{1}}\right\vert +\cdots+\left\vert \alpha_{z_{t}%
}\right\vert }{\sqrt{t}}\geq\sqrt{\frac{k}{N\left\lceil \log_{2}N\right\rceil
}}=\sqrt{\frac{k}{Nn}}.
\]
Here we can assume that $\alpha_{z_{1}},\ldots,\alpha_{z_{t}}$ are all
nonzero, since otherwise we simply decrease $t$. \ Now let $\beta_{z}$\ be the
element of $\left\{  1,-1,i,-i\right\}  $\ that is closest to $\alpha
_{z}/\left\vert \alpha_{z}\right\vert $, with ties broken arbitrarily. \ Then
our approximation to $\left\vert \psi\right\rangle $ will be the following:%
\[
\left\vert \phi\right\rangle :=\frac{1}{\sqrt{t}}\sum_{i=1}^{t}\beta_{z_{i}%
}\left\vert z_{i}\right\rangle .
\]
To specify $\left\vert \phi\right\rangle $,\ the classical advice just needs
to list $z_{1},\ldots,z_{t}$\ and $\beta_{z_{1}},\ldots,\beta_{z_{t}}$.
\ Since $t\leq k$, this requires at most $k\left(  n+2\right)  \leq m$\ bits.
\ Given the specification, it is clear that $\left\vert \phi\right\rangle
$\ can be prepared in time polynomial in $tn\leq m$. \ Moreover,%
\[
\left\langle \phi|\psi\right\rangle =\frac{1}{\sqrt{t}}\sum_{i=1}^{t}%
\beta_{z_{i}}^{\ast}\alpha_{z_{i}}\geq\frac{1}{\sqrt{t}}\sum_{i=1}^{t}%
\frac{\left\vert \alpha_{z_{i}}\right\vert }{\sqrt{2}}\geq\sqrt{\frac{k}{2Nn}%
}.
\]
We can therefore set $h:=\sqrt{\frac{k}{2Nn}}$,\ so that $k=2h^{2}Nn$. \ Hence%
\[
m\leq\left(  n+2\right)  \left(  k+1\right)  =\left(  n+2\right)  \left(
2h^{2}Nn+1\right)  =n+2+O\left(  h^{2}2^{n}n^{2}\right)  .
\]

\end{proof}

The following is an immediate consequence of Theorem \ref{explicit}.

\begin{corollary}
\label{explicitcor}Suppose we have an $n$-qubit quantum oracle $U$ such that
either (i) $U=U_{\psi}$\ for some $\left\vert \psi\right\rangle $, or (ii)
$U=I$\ is the identity. \ Then given an $m$-bit classical witness in support
of case (i), there exists a quantum algorithm that verifies the witness using
$O\left(  n\sqrt{2^{n}/m}+1\right)  $\ queries to $U$, together with $O\left(
n^{2}\sqrt{2^{n}/m}+\operatorname*{poly}\left(  m\right)  \right)  $\ steps of
auxiliary computation.
\end{corollary}

It is natural to ask whether we could construct a smaller explicit $h$-net,
and thereby improve the query complexity in Corollary \ref{explicitcor}\ from
$O\left(  n\sqrt{2^{n}/m}+1\right)  $\ to the optimal $O\left(  \sqrt{2^{n}
/m}+1\right)  $. \ We certainly believe that this is possible, but it seems to
require more complicated techniques from the theory of sphere coverings.

\section{Group Non-Membership\label{GNM}}

The Group Non-Membership ($\operatorname*{GNM}$) problem is defined as
follows. \ We are given a finite group $G$, a subgroup $H\leq G$, and an
element $x\in G$. \ The problem is to decide whether $x\notin H$.

But how are $G$, $H$, and $x$ specified? \ To abstract away the details of
this question, we will use Babai and Szemer\'{e}di's model of
\textit{black-box groups} \cite{bs:matrix}. \ In this model, we know
generators for $H$,\ and we know how to multiply and invert the elements of
$G$, but we \textquotedblleft do not know anything else.\textquotedblright%
\ \ More formally, we are given access to a group oracle $\mathcal{O}$, which
represents each element $x\in G$ by a randomly-chosen label $\ell\left(
x\right)  \in\left\{  0,1\right\}  ^{n}$\ for some $n\gg\log_{2}\left\vert
G\right\vert $. \ We are also given the labels of generators $\left\langle
h_{1},\ldots,h_{l}\right\rangle $\ for $H$. \ We are promised that every
element has a unique label.

Suppose that our quantum computer's state has the form%
\[
\left\vert \Phi\right\rangle =\sum_{x,y\in G,~z}\alpha_{x,y,z}\left\vert
\ell\left(  x\right)  ,\ell\left(  y\right)  \right\rangle \left\vert
z\right\rangle ,
\]
where $\ell\left(  x\right)  $\ and $\ell\left(  y\right)  $\ are labels of
group elements\ and $\left\vert z\right\rangle $\ is a workspace register.
\ Then the oracle $\mathcal{O}$\ maps this state to%

\[
\mathcal{O}\left\vert \Phi\right\rangle =\sum_{x,y\in G,~z}\alpha
_{x,y,z}\left\vert \ell\left(  x\right)  ,\ell\left(  xy^{-1}\right)
\right\rangle \left\vert z\right\rangle .
\]
Note that if the first register does not contain valid labels of group
elements, then $\mathcal{O}$\ can behave arbitrarily. \ Thus, from now on we
will ignore labels, and talk directly about the group elements they represent.
\ Using $\mathcal{O}$, it is easy to see that we can perform group inversion
(by putting the identity element $e$ in the $x$ register) and multiplication
(by first inverting $y$, then putting $y^{-1}$\ in the $y$ register), as well
as any combination of these operations.

We will show that $\operatorname*{GNM}$ has
polynomially-bounded\ $\mathsf{QCMA}$\ query complexity. \ In other words, if
$x\notin H$, then Merlin can provide Arthur with a $\operatorname*{poly}
\left(  n\right)  $-bit classical witness of that fact, which enables Arthur
to verify it with high probability using $\operatorname*{poly}\left(
n\right)  $\ quantum queries to the group oracle $\mathcal{O}$.

To prove this result, we first need to collect various facts from finite group
theory. \ Call $g_{1},\ldots,g_{k}$\ an \textit{efficient generating set} for
a finite group $G$ if (i) $k=O\left(  \log\left\vert G\right\vert \right)  $,
and (ii) every $x\in G$\ is expressible as $g_{1}^{e_{1}}\cdots g_{k}^{e_{k}}
$\ where $e_{1},\ldots,e_{k}\in\left\{  0,1\right\}  $. \ The following lemma
was shown by Babai and Erd\H{o}s \cite{be}.

\begin{lemma}
[\cite{be}]\label{reach}Every finite group $G$ has an efficient generating set.
\end{lemma}

Given finite groups $\Gamma$\ and $G$, we say that functions $f,g:\Gamma
\rightarrow G$\ are\ $\varepsilon$\textit{-close} if%
\[
\Pr_{x\in\Gamma}\left[  f\left(  x\right)  \neq g\left(  x\right)  \right]
\leq\varepsilon.
\]
Also, recall that $f:\Gamma\rightarrow G$\ is a homomorphism if $f\left(
xy\right)  =f\left(  x\right)  f\left(  y\right)  $\ for all $x,y\in\Gamma$.
\ The following two propositions relate $\varepsilon$-closeness to
homomorphisms.

\begin{proposition}
\label{homclose}If two homomorphisms $f,g:\Gamma\rightarrow G$\ are $\left(
1/2-\varepsilon\right)  $-close for any $\varepsilon>0$, then $f=g$.
\end{proposition}

\begin{proof}
Fix $x\in\Gamma$; then\ for all $y\in\Gamma$, we have $f\left(  x\right)
=f\left(  y\right)  f\left(  y^{-1}x\right)  $\ and $g\left(  x\right)
=g\left(  y\right)  g\left(  y^{-1}x\right)  $. \ By the union bound,%
\[
\Pr_{y\in\Gamma}\left[  f\left(  y\right)  =g\left(  y\right)  \wedge f\left(
y^{-1}x\right)  =g\left(  y^{-1}x\right)  \right]  \geq1-\Pr_{y\in\Gamma
}\left[  f\left(  y\right)  \neq g\left(  y\right)  \right]  -\Pr_{y\in\Gamma
}\left[  f\left(  y^{-1}x\right)  \neq g\left(  y^{-1}x\right)  \right]  >0.
\]
Hence there exists a $y$ such that $f\left(  y\right)  =g\left(  y\right)
$\ and $f\left(  y^{-1}x\right)  =g\left(  y^{-1}x\right)  $. \ But this
implies that $f\left(  x\right)  =g\left(  x\right)  $.
\end{proof}

In particular, Proposition \ref{homclose}\ implies that if a function $f$\ is
$1/5$-close\ to a homomorphism, then it is $1/5$-close\ to a \textit{unique}
homomorphism ($1/5$ being an arbitrary constant less than $1/4$).

\begin{proposition}
[Ben-Or et al. \cite{bclr}]\label{bclrprop}Given finite groups $\Gamma$\ and
$G$, a function $f:\Gamma\rightarrow G$, and a real number $\varepsilon>0$, if%
\[
\Pr_{x,y\in\Gamma}\left[  f\left(  xy\right)  \neq f\left(  x\right)  f\left(
y\right)  \right]  \leq\varepsilon
\]
then $f$ is $\varepsilon$-close\ to a homomorphism.
\end{proposition}

Together, Propositions \ref{homclose}\ and \ref{bclrprop} have the following
easy corollary.

\begin{corollary}
\label{homtest}There is a randomized algorithm which, given finite groups
$\Gamma$\ and $G$ and a function $f:\Gamma\rightarrow G$ as input, makes
$O\left(1\right)$ oracle queries to $f$, accepts with probability $1$ if $f$ is
a homomorphism, and rejects with probability at least $2/3$\ if $f$ is not
$1/5$-close to a homomorphism. \ Also, if $f$ is $1/5$-close to some
homomorphism $\widetilde{f}$, then there exists a randomized algorithm that,
given an input $x\in\Gamma$, makes $O\left(r\right)$\ oracle queries to $f$, and
outputs $\widetilde{f}\left(x\right)$\ with probability at least $1-1/2^{r}$.
\end{corollary}

In the present context, our algorithms are not limited in space or time, and we
can say for simplicity that $\Gamma$ is represented by its entire multiplication
table.  It is then easy, as the proof will require, to pick elements of $\Gamma$
uniformly at random.  By contrast, $G$ is represented by oracle access, but
there will be no need to choose its elements at random.

\begin{proof}
The first algorithm simply chooses $O\left(  1\right)  $\ pairs $x,y\in\Gamma
$\ uniformly at random, accepts if $f\left(  xy\right)  =f\left(  x\right)
f\left(  y\right)  $\ for all of them, and rejects otherwise. \ Let
$k=O\left(  r\right)  $. \ Then the second algorithm chooses $z_{1},
\ldots,z_{k}\in\Gamma$\ uniformly at random, and outputs the plurality answer
among $f\left(  z_{1}\right)  f\left(  z_{1}^{-1}x\right)  ,\ldots,f\left(
z_{k}\right)  f\left(  z_{k}^{-1}x\right)  $\ (breaking ties arbitrarily).
\end{proof}

It follows from the Classification of Finite Simple Groups that there are at
most two finite simple groups of any particular order (see \cite{atlas}\ for
example). \ The following well-known result is a combination of that fact and
of a theorem due to Neumann \cite{neumann}.

\begin{theorem}
\label{iso}There are $N^{O\left(\left(\log_{2}N\right)^2\right)}$ groups of
order $N$ up to isomorphism.\footnote{The most accurate asymptotic result on
the number of groups of order $N$, in terms of the prime factorization of $N$,
appears in a paper by Pyber \cite{pyber}.}
\end{theorem}

Finally, recall that the Hidden Subgroup Problem ($\operatorname*{HSP}$) is
defined as follows. \ We are given a finite group $G$, and oracle access to a
function $f:G\rightarrow\mathbb{Z}$. \ We are promised that there exists a
\textquotedblleft hidden subgroup\textquotedblright\ $H\leq G$\ such that
$f\left(  x\right)  =f\left(  y\right)  $\ if and only if $x$\ and $y$ belong
to the same left coset of $H$.\ \ The problem is then to output a set of
generators for $H$. \ Whether $\operatorname*{HSP}$\ can be solved in quantum
polynomial time, for various non-abelian groups $G$, is one of the most
actively studied questions in quantum computing. \ However, if we only care
about query complexity, then Ettinger, H\o yer, and Knill \cite{ehk} proved
the following useful result.

\begin{theorem}
[\cite{ehk}]\label{ehkthm}There is a quantum algorithm such that, given any
finite group $G$ as oracular input, solves $\operatorname*{HSP}$ using only
$\operatorname*{polylog}\left(  \left\vert G\right\vert \right)  $\ quantum
queries to $f$ (together with a possibly exponential amount of
postprocessing).\footnote{Indeed, for \textit{Normal} $\operatorname*{HSP}$
(which is the special case we care about), Hallgren, Russell, and Ta-Shma
\cite{hrt}\ improved this result, showing how to find a hidden subgroup\ using
only $O\left(\log\left\vert G\right\vert\right)$ queries to $f$ (again,
with exponential postprocessing).}
\end{theorem}

We can now prove Theorem \ref{hspthm}: that $\operatorname*{GNM}$ has
polynomially-bounded $\mathsf{QCMA}$\ query complexity.

\begin{proof}
[Proof of Theorem \ref{hspthm}]Let $G$\ be a group of order at most $2^{n}$,
and let $\mathcal{O}$ be a group oracle that maps each element of $G$ to an
$n$-bit label. \ Also, given (the labels of) group elements $x,h_{1},
\ldots,h_{m}\in G$, let $H$\ be the subgroup of $G$ generated by
$\left\langle h_{1},\ldots,h_{m}\right\rangle $. \ Then the problem is to
decide if $x\notin H$.

In our $\mathsf{QCMA}$\ protocol for this problem, Merlin's witness will
consist of the following:

\begin{itemize}
\item An explicit \textquotedblleft model group\textquotedblright\ $\Gamma$,
of order at most $2^{n}$.
\item A list of elements $\gamma_{1},\ldots,\gamma_{k}\in\Gamma$, where
$k=O\left(\log\left\vert \Gamma\right\vert \right)$.
\item A corresponding list $g_{1},\ldots,g_{k}\in G$.
\item Another list $z,\lambda_{1},\ldots,\lambda_{m}\in\Gamma$.
\end{itemize}

We should be more explicit about the notion of an ``explicit'' group $\Gamma$,
and about the syntax of this witness. By Theorem \ref{iso}, there are at most
$2^{\operatorname*{poly}\left( n\right)}$\ groups of order $\left\vert
\Gamma\right\vert \leq2^{n}$\ up to isomorphism.  Since Arthur is allowed
unlimited computation and is only  restricted in queries, he can construct a
full multiplication table for $\Gamma$ using only the name of its isomorphism
type.  The multiplication table is not unique, because the elements of $\Gamma$
can be permuted; but for instance Arthur could construct the lexicographically
first such table.  Since Merlin can anticipate Arthur's construction of
$\Gamma$, he can then refer to elements of $\Gamma$ using the same construction.
He can also refer to elements of $G$ since he understands the oracle. In
conclusion, Merlin can specify the witness using only
$\operatorname*{poly}\left(n\right)$ bits.

If Merlin is honest, then the witness will have the following three
properties:

\begin{enumerate}
\item[(1)] $\gamma_{1},\ldots,\gamma_{k}$ is an efficient generating set for
$\Gamma$.

\item[(2)] $z\notin\Lambda$, where $\Lambda$\ is the subgroup of $\Gamma
$\ generated by $\left\langle \lambda_{1},\ldots,\lambda_{m}\right\rangle $.

\item[(3)] There exists an embedding $\widetilde{f}:\Gamma\rightarrow G$, such
that (i) $\widetilde{f}\left(  \gamma_{i}\right)  =g_{i}$\ for all
$i\in\left\{  1,\ldots,k\right\}  $, (ii) $\widetilde{f}\left(  \lambda
_{j}\right)  =h_{j}$ for all $j\in\left\{  1,\ldots,m\right\}  $, and (iii)
$\widetilde{f}\left(  z\right)  =x$.
\end{enumerate}

Suppose for the moment that (1)-(3) all hold. \ Then there exists an embedding
$\widetilde{f}:\Gamma\rightarrow G$, which maps the set $\left\langle
\gamma_{1},\ldots,\gamma_{k}\right\rangle $\ in $\Gamma$\ to the set
$\left\langle g_{1},\ldots,g_{k}\right\rangle $ in $G$. \ Furthermore, this
embedding satisfies $\widetilde{f}\left(  \Lambda\right)  =H$\ and
$\widetilde{f}\left(  z\right)  =x$. \ Since $z\notin\Lambda$\ by (2), it
follows that $x\notin H$ as well, which is what Arthur wanted to check.

So it suffices to verify (1)-(3). \ In the remainder of the proof, we will
explain how to do this using a possibly exponential amount of computation, but
only $\operatorname*{poly}\left(  n\right)  $\ quantum queries to the group
oracle $\mathcal{O}$.

First, since properties (1) and (2) only involve the explicit group $\Gamma$,
not the black-box group $G$, Arthur can verify these properties
\textquotedblleft free of cost.\textquotedblright\ \ In other words,
regardless of how much computation he needs, he never has to query the group
oracle.

The nontrivial part is to verify (3). \ It will be convenient to split (3)
into the following sub-claims:

\begin{enumerate}
\item[(3a)] There exists a homomorphism $\widetilde{f}:\Gamma\rightarrow G$
such that $\widetilde{f}\left(  \gamma_{i}\right)  =g_{i}$\ for all
$i\in\left\{  1,\ldots,k\right\}  $.

\item[(3b)] $\widetilde{f}$\ satisfies $\widetilde{f}\left(  z\right)
=x$\ and $\widetilde{f}\left(  \lambda_{j}\right)  =h_{j}$ for all
$j\in\left\{  1,\ldots,m\right\}  $.

\item[(3c)] $\widetilde{f}$\ is injective (i.e. is an embedding into $G$).
\end{enumerate}

To verify (3a), first Arthur fixes a \textquotedblleft canonical
representation\textquotedblright\ of each element $\gamma\in\Gamma$. \ This
representation has the form%
\[
\gamma=\gamma_{1}^{e_{1}}\cdots\gamma_{k}^{e_{k}},
\]
where $\left\langle \gamma_{1},\ldots,\gamma_{k}\right\rangle $ is the
efficient generating set for $\Gamma$, and $e_{1},\ldots,e_{k}\in\left\{
0,1\right\}  $ are bits depending on $\gamma$.\ \ Next he defines a function
$f:\Gamma\rightarrow G$\ by%
\[
f\left(  \gamma\right)  :=g_{1}^{e_{1}}\cdots g_{k}^{e_{k}}%
\]
for all $\gamma\in\Gamma$. \ By using the canonical representation of $\gamma
$, Arthur can evaluate $f\left(  \gamma\right)  $\ using at most $k-1$ queries
to the group oracle $\mathcal{O}$. \ Finally Arthur appeals to Corollary
\ref{homtest}. \ If $f$ is not $1/5$-close\ to a homomorphism, then by using
$O\left(  1\right)  $\ queries to $f$, with high probability Arthur can detect
that $f$ is not a homomorphism. \ In that case Merlin has been caught
cheating, so Arthur rejects. \ On the other hand, if $f$\ is $1/5$-close to
some homomorphism $\widetilde{f}$, then by using $O\left(  \log\left\vert
\Gamma\right\vert \right)  $\ queries to $f$, with high probability Arthur can
\textquotedblleft correct\textquotedblright\ $f$ to $\widetilde{f}$. \ In that
case it remains only to check that $\widetilde{f}\left(  \gamma_{i}\right)
=g_{i}$\ for all $i\in\left\{  1,\ldots,k\right\}  $.

Once Arthur has an efficient procedure for computing $\widetilde{f}$---that
is, a procedure that involves only $\operatorname*{poly}\left(  n\right)
$\ queries to $\mathcal{O}$---he can then verify property (3b) directly.

To verify (3c), Arthur runs the algorithm of Ettinger, H\o yer, and Knill
\cite{ehk} for the Hidden Subgroup Problem. \ Notice that, since
$\widetilde{f}:\Gamma\rightarrow G$\ is a homomorphism, there must be a
\textquotedblleft hidden subgroup\textquotedblright\ $K\leq\Gamma$---namely
the kernel of $\widetilde{f}$---such that $\widetilde{f}$\ is constant on
cosets of $K$\ and distinct on distinct cosets. \ Furthermore, $\widetilde{f}$\ is injective if and only if $K$ is trivial. \ But deciding whether $K$ is
trivial is just an instance of $\operatorname*{HSP}$, and can therefore be
solved using $\operatorname*{poly}\left(  n\right)  $\ quantum queries\ by
Theorem \ref{ehkthm}.
\end{proof}

\subsection{Computational Complexity\label{COMP}}

Theorem\ \ref{hspthm}\ showed that one can always verify group non-membership
using a polynomial-size classical witness, together with polynomially many
quantum queries to the group oracle $\mathcal{O}$. \ Unfortunately, while the
\textit{query} complexity is polynomial, the \textit{computational} complexity
might be exponential. \ However, as mentioned in Section \ref{RESULTS}, we
conjecture that this shortcoming of Theorem \ref{hspthm} can be removed, and
that $\operatorname*{GNM}$ is in $\mathsf{QCMA}$\ for any group oracle
$\mathcal{O}$.

In our $\mathsf{QCMA}$ protocol, the main computational problem that needs to be
solved is not the general $\operatorname*{HSP}$, but rather the Normal Hidden
Subgroup Problem ($\operatorname*{NHSP}$)---that is, $\operatorname*{HSP}$\
where the hidden subgroup is normal. \ This is because the kernel of a
homomorphism is always a normal subgroup. \ Hallgren, Russell, and Ta-Shma
\cite{hrt} showed that $\operatorname*{NHSP}$\ is in $\mathsf{BQP}$ for an
explicit group $\Gamma$, provided that the quantum Fourier transform over\
$\Gamma$ can be implemented efficiently (and its output can  be interpreted). \
Furthermore, Moore, Rockmore, and Russell \cite{mrr} showed that many classes of
finite groups $G$ have an explicit model $\Gamma\cong G$\ for which this
assumption holds.

Even if $\operatorname*{NHSP}$\ is in $\mathsf{BQP}$, there are two remaining
obstacles to showing that $\operatorname*{GNM}$ is in $\mathsf{QCMA}$. \ First,
we need to be able to verify group non-membership in the explicit model group
$\Gamma$, possibly with the help of additional classical information from
Merlin. \ Second, we need an efficient algorithm to compute the function
$\widetilde{f}:\Gamma\rightarrow G$ for every $\gamma\in\Gamma$, even though
$\widetilde{f}$ is explicitly defined only on the generators
$\gamma_1,\ldots,\gamma_k$.

In the context of computational complexity (as opposed to query complexity), the
notion of an ``explicit group" needs to be better explained. Keeping in mind
that this entire section is only one possible path to showing that
$\operatorname*{GNM}$ is in $\mathsf{QCMA}$, here is one definition that
captures the ideas of previous sections.

\begin{definition} A (polylog-time) \emph{explicit sequence} of finite groups is
a sequence $\Gamma_n$ such that each term is a group law on the set
$\{1,\ldots,|\Gamma_n|\}$.  Moreover the multiplication function $m(n,x,y) = xy$
and the inversion function $i(n,x) = x^{-1}$ can both be computed in polynomial
time in $\log |\Gamma_n|$.  An explicit sequence is \emph{universal} if every
finite group is isomorphic to at least one $\Gamma_n$.
\end{definition}

For example, the symmetric group (sequence) $S_n$ is explicit, because the
standard notation for permutations can be compressed to the integers from 1 to
$n!$.  Likewise the matrix groups $\mathrm{GL}(n,q)$ form an explicit sequence
in the joint parameter $(n,a(x))$, where $a(x)$ is a polynomial whose splitting
field is $\mathbb{F}_q$.  But there is no reason to believe that an explicit
model of a group is unique up to polylog-time bijections.  On the contrary, if
the discrete logarithm problem is hard, then $(\mathbb{Z}/q)^\times$ and
$\mathbb{Z}/(q-1)$ are inequivalent explicit models for isomorphic groups.

It is not known whether there is a universal explicit sequence of finite
groups.  The current best result for solvable groups is quasipolylogarithmic
time \cite{hofling}.  Theorem~\ref{iso} implies that the number of isomorphism
classes of finite groups does not by itself preclude a universal explicit
sequence.

If there is a universal sequence of explicit finite groups with the following
additional properties, then following the methods of the previous section, it
would show that $\operatorname*{GNM}$ is in $\mathsf{QCMA}$. (We drop the formal
subscript $n$.)


\begin{enumerate}
\item[(i)] Each $\Gamma$ has a list of generators
$\gamma_1,\ldots,\gamma_k\in\Gamma$ that can be computed in
$O(\operatorname*{polylog}|\Gamma|)$ time.  Moreover, given an element
$\gamma\in\Gamma$, there is a polylog algorithm to express it as a
(polylogarithmic length) product of the generators. It suffices if
this algorithm is polylog time on average for random $\gamma$.  A straight-line
program rather than a product also suffices.

\item[(ii)] $\operatorname*{NHSP}$ over $\Gamma$\ is in $\mathsf{BQP}$. 

\item[(iii)] $\operatorname*{GNM}$ over $\Gamma$ is in $\mathsf{QCMA}$.
\end{enumerate}

If the Hallgren-Russell-Ta-Shma algorithm is used for condition (iii), then
there should be an algorithm for the quantum Fourier transform over $\Gamma$. 
In this case the QFT produces randomly-chosen characters of the quotient group
$\Gamma/\Lambda$ for some normal subgroup $\Lambda$ (the hidden subgroup).  The
characters must also be listed in some explicit form so that be encoded so that
$\Lambda$ can be recognized in polylog time, or at least that the triviality of
$\Lambda$ can be so recognized. (Here too, ``explicit" means that the characters
of $\Gamma$ are numbered consecutively and that the relevant algorithms use this
numbering.)

If $\Gamma$ is the symmetric group $S_n$, or an abelian group expressed as a
product of cyclic groups, or if it is a matrix group
$\mathrm{GL}(n,\mathbb{Z}/q)$, then there is an easy generating set that
satisfies (i) and (ii) (exercise for the reader).   In the abelian case,
$\operatorname*{NHSP}$ is in $\mathsf{BQP}$ by the work of Shor \cite{shor}\ and
Kitaev \cite{kitaev:meas}; $\operatorname*{GNM}$ is in $\mathsf{P}$ by linear
algebra.  If $\Gamma = S_n$, then $\operatorname*{NHSP}$ is trivial (since the
only normal subgroup is $A_n$) and $\operatorname*{GNM}$\ is in $\mathsf{P}$\ by
the work of Sims \cite{sims}.  Meanwhile Babai and Szemer\'{e}di
\cite{bs:matrix} showed that if every finite simple group has an explicit
polylogarithmic presentation, then $\operatorname*{GNM}$ is in $\mathsf{NP}$ for
$\mathrm{GL}(n,\mathbb{Z}/q)$.

Since the point of condition (ii) is to allow Arthur to confirm Merlin's claimed
homomorphism from $\Gamma$ to $G$, a polylogarithmic presentation of $\Gamma$
would yield an alternative method that does not rely on the algorithm of
Corollary \ref{homtest}.  The status of this problem is that the only unknown
case among finite simple groups is Ree groups of type ${}^2G_2(q)$; all other
finite simple groups are known to have such a presentation
\cite{bgklp,hulpke-seress}.  If two finite groups $\Gamma$ and $\Lambda$ have
polylog-length presentations, then so does an extension of $\Gamma$ by
$\Lambda$; but there is no known polylog-time algorithm to generate the
presentations of such extensions.  In summary, the groups ${}^2G_2(q)$ and the
extension problem are the remaining obstructions a universal, explicit sequence
of polylog presentations of finite groups, which would provide a simple
alternative to condition (ii).  Regardless, all known QFT algorithms employ
flags of subgroups, which are structures that can also be used to satisfy
condition (ii).

Obviously the entire program is far from complete, and each step is open to
variations.  But we optimistically conjecture that all steps can be completed
for arbitrary finite groups.

\section{Mimicking Random Quantum Oracles\label{SQO}}

We have seen, on the one hand, that there exists a quantum oracle separating
$\mathsf{QMA}$\ from $\mathsf{QCMA}$; and on the other hand, that separating
these classes by a \textit{classical} oracle seems much more
difficult.\ \ Together, these results raise a general question: how much
\textquotedblleft stronger\textquotedblright\ are quantum oracles than
classical ones?\ \ In particular, are there complexity classes $\mathcal{C}$
and $\mathcal{D}$ that can be separated by quantum oracles, but such that
separating them by classical oracles is almost as hard as separating them in
the unrelativized world? \ Whatever the answer, we conjecture that
$\mathsf{QMA}$\ and $\mathsf{QCMA}$\ are \textit{not} examples of such
classes. \ The reason is that it seems possible, using only classical oracles,
to approximate\ quantum oracles similar to ones that would separate
$\mathsf{QMA}$\ from $\mathsf{QCMA}$.

To illustrate, let $\sigma$\ be the uniform probability measure over
$2^{n}\times2^{n}$\ unitary diagonal matrices. \ (In other words, each
diagonal entry of $D\in\sigma$\ is a random complex number with norm $1$.)
\ Also, let $H^{\otimes n}$\ be a tensor product of $n$ Hadamard matrices.
\ Then let $\varsigma_{k}$\ be the probability measure over $2^{n}\times2^{n}%
$\ unitary matrices%
\[
U=D_{k}H^{\otimes n}D_{k-1}H^{\otimes n}\cdots H^{\otimes n}D_{1}H^{\otimes n}%
\]
induced by drawing each $D_{i}$\ independently from $\sigma$. \ In other
words, $U\in\varsigma_{k}$\ is obtained by first applying a Hadamard gate to
each qubit, then a random $2^{n}\times2^{n}$\ diagonal matrix, then Hadamard
gates again, then another random diagonal matrix, and so on $k$ times.

Note that we can efficiently apply such a $U$---at least to polynomially
many\ bits of precision---if given a classical random oracle $A$. \ To do so,
we simply implement the random diagonal matrix $D_{i}$\ as%
\[
\sum_{x\in\left\{  0,1\right\}  ^{n}}\alpha_{x}\left\vert x\right\rangle
\mapsto\sum_{x\in\left\{  0,1\right\}  ^{n}}\omega^{A\left(  i,x\right)
}\alpha_{x}\left\vert x\right\rangle ,
\]
where $A\left(  i,x\right)  $\ is a uniformly random $n$-bit integer indexed
by $i$ and $x$, and $\omega=e^{2\pi i/2^{n}}$.

Now\ let $\mu$\ be the uniform probability measure over $2^{n}\times2^{n}$
unitary matrices. \ If $k\ll2^{n}$, then $\varsigma_{k}$ is \textit{not} close
to $\mu$ in variation distance, since the former has only $\Theta\left(
k2^{n}\right)  $ degrees of freedom while the latter has $\Theta\left(
k4^{n}\right)  $.\footnote{Admittedly, it is still conceivable that the
finite-precision version of $\varsigma_{k}$\ is close in variation distance to
the finite-precision version of $\mu$. \ However, a more sophisticated
argument that counts distinguishable unitaries rules out that possibility as
well.} \ On the other hand, we conjecture that a $U$\ drawn from
$\varsigma_{k}$\ will \textquotedblleft look random\textquotedblright\ to any
polynomial-time algorithm, and that this property can be used to prove a
classical oracle separation between $\mathsf{QMA}$\ and $\mathsf{QCMA}$.

Let us explain what we mean in more detail. \ Suppose we are given access\ to
an $n$-qubit unitary oracle $U$, and want to decide whether

\begin{enumerate}
\item[(i)] $U$ was drawn uniformly at random (that is, from $\mu$), or

\item[(ii)] $U$ was drawn uniformly at random conditioned on there existing
$n/2$-qubit pure states $\left\vert \psi\right\rangle $\ and $\left\vert
\varphi\right\rangle $\ such that $U\left(  \left\vert 0\right\rangle
^{\otimes n/2}\left\vert \psi\right\rangle \right)  \approx\left\vert
0\right\rangle ^{\otimes n/2}\left\vert \varphi\right\rangle $.
\end{enumerate}

In case (i), the states $\left\vert \psi\right\rangle $\ and $\left\vert
\varphi\right\rangle $ will exist only with negligible
probability.\footnote{Indeed, the reason we did not ask for $\left(
n-1\right)  $-qubit states $\left\vert \psi\right\rangle $\ and $\left\vert
\varphi\right\rangle $\ such that $U\left(  \left\vert 0\right\rangle
\left\vert \psi\right\rangle \right)  \approx\left\vert 0\right\rangle
\left\vert \varphi\right\rangle $ is that such states will exist (almost)
generically. \ For the choice of $\left\vert \psi\right\rangle $\ gives us
$2^{n-1}-1$\ independent complex variables, whereas the requirement that
$U\left(  \left\vert 0\right\rangle \left\vert \psi\right\rangle \right)
$\ have the form $\left\vert 0\right\rangle \left\vert \varphi\right\rangle
$\ imposes only $2^{n-1}$\ constraints. \ Asking for $\left(  n-2\right)
$-qubit states $\left\vert \psi\right\rangle $\ and $\left\vert \varphi
\right\rangle $\ such that $U\left(  \left\vert 00\right\rangle \left\vert
\psi\right\rangle \right)  \approx\left\vert 00\right\rangle \left\vert
\varphi\right\rangle $\ might suffice (since now we have $2^{n-2}%
-1$\ variables versus $3\cdot2^{n-2}$\ constraints), but we wish to stay on
the safe side.} \ It follows that the above problem\ is in $\mathsf{QMA}^{U}%
$---since if case (ii) holds, then a succinct quantum proof of that fact is
just $\left\vert \psi\right\rangle $ itself. \ We now state three conjectures
about this problem, in increasing order of difficulty.

\begin{conjecture}
\label{notinqcma}The above problem is not in $\mathsf{QCMA}^{U}$. \ In other
words, if case (ii) holds, there is no succinct classical proof of that fact
that can be verified with high probability using $\operatorname*{poly}\left(
n\right)  $ quantum queries to $U$.
\end{conjecture}

Presumably Conjecture \ref{notinqcma}\ can be proved using ideas similar to
those in Section \ref{QORACLE}. \ If so, then the next step is to replace the
uniform measure $\mu$\ by the \textquotedblleft pseudorandom\textquotedblright%
\ measure $\varsigma_{k}$.

\begin{conjecture}
\label{secondconj}Suppose that instead of being drawn from $\mu$, the unitary
$U$ is drawn from $\varsigma_{k}$ for some $k=\Omega\left(  n\right)  $.
\ Then the probability that there exist $n/2$-qubit states $\left\vert
\psi\right\rangle $\ and $\left\vert \varphi\right\rangle $\ such that
$U\left(  \left\vert 0\right\rangle ^{\otimes n/2}\left\vert \psi\right\rangle
\right)  \approx\left\vert 0\right\rangle ^{\otimes n/2}\left\vert
\varphi\right\rangle $ is still negligibly small.
\end{conjecture}

Now suppose we want to decide whether

\begin{enumerate}
\item[(i')] $U$ was drawn from $\varsigma_{k}$, or

\item[(ii')] $U$ was drawn from $\varsigma_{k}$\ conditioned on there existing
$n/2$-qubit states $\left\vert \psi\right\rangle $\ and $\left\vert
\varphi\right\rangle $\ such that $U\left(  \left\vert 0\right\rangle
^{\otimes n/2}\left\vert \psi\right\rangle \right)  \approx\left\vert
0\right\rangle ^{\otimes n/2}\left\vert \varphi\right\rangle $.
\end{enumerate}

Also, let $A$ be a classical oracle that encodes the diagonal matrices
$D_{1},\ldots,D_{k}$\ such that%
\[
U=D_{k}H^{\otimes n}D_{k-1}H^{\otimes n}\cdots H^{\otimes n}D_{1}H^{\otimes
n}.
\]
If Conjecture \ref{secondconj} is true, then case (ii') can be verified in
$\mathsf{QMA}^{A}$. \ So to obtain a classical oracle separation between
$\mathsf{QMA}$ and $\mathsf{QCMA}$, the one remaining step would be to prove
the following.

\begin{conjecture}
\label{thirdconj}Case (ii') cannot be verified in $\mathsf{QCMA}^{A}$.
\end{conjecture}

\subsection{From Random Oracles to Random Unitaries\label{RORU}}

The previous discussion immediately suggests even simpler questions about the
ability of classical oracles to mimic quantum ones. \ In particular, could a
$\mathsf{BQP}$\ machine\ use a classical random oracle to prepare a uniformly
random $n$-qubit pure state? \ Also, could it use such an oracle to apply a
random $n$-qubit unitary?

In this section we answer the first question in the affirmative, and present
partial results about the second question. \ We first need a notion that we
call the \textquotedblleft$\varepsilon$-smoothing\textquotedblright\ of a
probability measure.

\begin{definition}
\label{esmooth}Let $\sigma$\ be a probability measure over $\left\vert
\psi\right\rangle \in\mathbb{CP}^{2^{n}-1}$. \ Then the $\varepsilon
$-smoothing of $\sigma$, or $\mathcal{S}_{\varepsilon}\left(  \sigma\right)
$, is the probability measure obtained by first drawing a state $\left\vert
\psi\right\rangle $\ from $\sigma$, and then drawing a state $\left\vert
\varphi\right\rangle $\ uniformly at random subject to $\left\langle
\varphi|\psi\right\rangle \geq1-\varepsilon$.
\end{definition}

Let $\mu$\ be the uniform measure over $\mathbb{CP}^{2^{n}-1}$. \ Also, let
$Q$ be a quantum algorithm that queries a classical oracle $A$. \ Suppose
that, given $0^{n}$\ as input, $Q^{A}$\ outputs the pure state $\left\vert
\psi_{A}\right\rangle \in\mathbb{CP}^{2^{n}-1}$. \ Then we say that
$Q$\ \textquotedblleft approximates the uniform measure within $\varepsilon
$\textquotedblright\ if, as we range over uniform random\ $A\subseteq\left\{
0,1\right\}  ^{n}$, the induced probability measure $\sigma$\ over $\left\vert
\psi_{A}\right\rangle $\ satisfies $\left\Vert \mathcal{S}_{\varepsilon
}\left(  \sigma\right)  -\mu\right\Vert \leq\varepsilon$.

\begin{theorem}
\label{randstate}For all polynomials $p$, there exists a quantum algorithm $Q$
that runs in polynomial time, and that approximates the uniform
measure\ within $2^{-p\left(  n\right)  }$.
\end{theorem}

\begin{proof}
[Proof Sketch]The algorithm $Q$ is as follows: first prepare a uniform
superposition over $n$-bit strings. \ Then, using the classical random oracle
$A$ as a source of random bits, map this state to%
\[
\left\vert \Psi\right\rangle =\frac{1}{2^{n/2}}\sum_{x\in\left\{  0,1\right\}
^{n}}\left\vert x\right\rangle \left(  \sqrt{1-\left\vert \alpha
_{x}\right\vert ^{2}}\left\vert 0\right\rangle +\alpha_{x}\left\vert
1\right\rangle \right)  ,
\]
where each $\alpha_{x}$ is essentially a Gaussian random variable. \ More
precisely, let $q\left(  n\right)  =\left(  n+p\left(  n\right)  \right)
^{2}$. \ Then each $\alpha_{x}$\ is drawn independently from a complex
Gaussian distribution with mean $0$ and variance $1/q\left(  n\right)  $, with
the two technicalities that (1) $\alpha_{x}$\ is rounded to $q\left(
n\right)  $\ bits of precision, and (2) the cutoff $\left\vert \alpha
_{x}\right\vert \leq1$\ is imposed. \ (By a tail bound, with overwhelming
probability we will have $\left\vert \alpha_{x}\right\vert \leq1$\ for all $x$
anyway.)

Next measure the second register of $\left\vert \Psi\right\rangle $\ in the
standard basis. The outcome $\left\vert 1\right\rangle $\ will be observed
with probability $\Omega\left(  1/q\left(  n\right)  \right)  $.
\ Furthermore, conditioned on $\left\vert 1\right\rangle $ being observed, one
can check that the distribution $\sigma$\ over the reduced state of the first
register satisfies $\left\Vert \mathcal{S}_{2^{-p\left(  n\right)  }}\left(
\sigma\right)  -\mu\right\Vert \leq2^{-p\left(  n\right)  }$. \ (We omit the
calculation.) \ Hence it suffices to repeat the algorithm $O\left(  q\left(
n\right)  \right)  $\ times.
\end{proof}

Theorem \ref{randstate}\ shows that, by using a classical random oracle $A$,
we can efficiently prepare a uniformly random $n$-qubit state $\left\vert
\psi_{A}\right\rangle $. \ But what if we want to use a random oracle to apply
a uniformly random $n$-qubit \textit{unitary} $U_{A}$? \ It is clear that we
can do this if we have exponential time:\ given an oracle $A$, we simply query
an exponentially long prefix $A^{\ast}$\ of $A$, and then treat $A^{\ast}$\ as
an explicit description of a quantum circuit for $U_{A}$.\ \ But what if we
can make only polynomially many quantum queries to $A$? \ We do not know
whether that suffices for applying a random unitary; indeed, we do not even
have a conjecture about this.

What we \textit{can} show is that a single quantum query to the classical
oracle $A$ does not suffice for applying a random unitary. \ In particular,
suppose every entry of an $n$-qubit unitary matrix $U_{A}$\ is a degree-$1$
polynomial in the bits of $A$ (as it must be, if $U_{A}$\ is the result of a
single quantum query). \ Then $U_{A}$\ can assume at most $4^{2^{n}}%
$\ distinct values as we range over the possible $A$'s, as opposed to the
$\Omega\left(  c^{2^{2n}}\right)  $\ that would be needed to approximate every
$n$-qubit unitary. \ To prove this statement, we first need a lemma about
matrices satisfying a certain algebraic relation.

\begin{lemma}
\label{matrixlem}Let $E_{1},\ldots,E_{M}$\ be nonzero $N\times N$\ matrices
over $\mathbb{C}$, and suppose that $E_{i}E_{j}^{\dag}+E_{j}E_{i}^{\dag}%
=0$\ for all $i\neq j$. \ Then $M\leq2N$.
\end{lemma}

\begin{proof}
Suppose by contradiction that $M>2N$. \ Let $e_{i}^{\left(  k\right)  }$\ be
vector in $\mathbb{C}^{N}$\ corresponding to the $k^{th}$\ row of $E_{i}$.
\ Then the condition $E_{i}E_{j}^{\dag}+E_{j}E_{i}^{\dag}=0$\ implies that%
\[
e_{i}^{\left(  k\right)  }\cdot e_{j}^{\left(  l\right)  }+e_{j}^{\left(
k\right)  }\cdot e_{i}^{\left(  l\right)  }=0
\]
for all $i\neq j$\ and $k,l$, where $\cdot$\ denotes the complex inner
product. \ Now for all $i$, let $k\left(  i\right)  $\ be the minimum $k$ such
that $e_{i}^{\left(  k\right)  }\neq0$, and consider the vectors
$e_{1}^{\left(  k\left(  1\right)  \right)  },\ldots,e_{M}^{\left(  k\left(
M\right)  \right)  }\in\mathbb{C}^{N}$. \ Certainly these vectors are not all
orthogonal---indeed, since $M>2N$, there must exist $i\neq j$\ such that
$\operatorname{Re}\left(  e_{i}^{\left(  k\left(  i\right)  \right)  }\cdot
e_{j}^{\left(  k\left(  j\right)  \right)  }\right)  \neq0$. \ There are now
two cases: if $k\left(  i\right)  =k\left(  j\right)  $, then%
\[
e_{i}^{\left(  k\left(  i\right)  \right)  }\cdot e_{j}^{\left(  k\left(
i\right)  \right)  }+e_{j}^{\left(  k\left(  i\right)  \right)  }\cdot
e_{i}^{\left(  k\left(  i\right)  \right)  }\neq0
\]
and we are done. \ On the other hand, if $k\left(  i\right)  \neq k\left(
j\right)  $, then%
\[
e_{j}^{\left(  k\left(  i\right)  \right)  }\cdot e_{i}^{\left(  k\left(
j\right)  \right)  }=-e_{i}^{\left(  k\left(  i\right)  \right)  }\cdot
e_{j}^{\left(  k\left(  j\right)  \right)  }%
\]
is nonzero. \ Hence $e_{j}^{\left(  k\left(  i\right)  \right)  }$\ and
$e_{i}^{\left(  k\left(  j\right)  \right)  }$\ must themselves be nonzero.
\ But if $k\left(  i\right)  >k\left(  j\right)  $, then this contradicts the
minimality of $k\left(  i\right)  $, while if $k\left(  i\right)  <k\left(
j\right)  $\ then it contradicts the minimality of $k\left(  j\right)  $.
\end{proof}

We can now prove the main result.

\begin{theorem}
\label{onequery}Let $U\left(  X\right)  $\ be an $N\times N$ matrix, every
entry of which is a degree-$1$ complex polynomial in variables $X=\left(
x_{1},\ldots,x_{k}\right)  $. \ Suppose $U\left(  X\right)  $\ is unitary for
all $X\in\left\{  0,1\right\}  ^{k}$. \ Then $U\left(  X\right)  $ can assume
at most $4^{N}$\ distinct values\ as we range over $X\in\left\{  0,1\right\}
^{k}$.
\end{theorem}

\begin{proof}
By suitable rotation, we can assume without loss of generality that $U\left(
0^{k}\right)  $\ is the $N\times N$\ identity $I$. \ Let $X_{i}$\ be the
$k$-bit string with a `$1$' only in the $i^{th}$\ position,\ and let
$E_{i}:=U\left(  X_{i}\right)  -I$. \ Then\ for all $i$,%
\begin{align*}
E_{i}E_{i}^{\dag}  &  =\left(  U\left(  X_{i}\right)  -I\right)  \left(
U\left(  X_{i}\right)  ^{\dag}-I^{\dag}\right) \\
&  =I-U\left(  X_{i}\right)  -U\left(  X_{i}\right)  ^{\dag}+I\\
&  =-E_{i}-E_{i}^{\dag}.
\end{align*}
Next, for all $i\neq j$, let $X_{ij}$\ be the $k$-bit string with `$1$'s\ only
in the $i^{th}$\ and $j^{th}$\ positions. \ Since $U\left(  X\right)  $\ is an
affine function of $X$, we have%
\begin{align*}
U\left(  X_{ij}\right)   &  =U\left(  0^{k}\right)  +\left(  U\left(
X_{i}\right)  -U\left(  0^{k}\right)  \right)  +\left(  U\left(  X_{j}\right)
-U\left(  0^{k}\right)  \right) \\
&  =I+E_{i}+E_{j}.
\end{align*}
Therefore%
\begin{align*}
0  &  =U\left(  X_{ij}\right)  U\left(  X_{ij}\right)  ^{\dag}-I\\
&  =\left(  I+E_{i}+E_{j}\right)  \left(  I^{\dag}+E_{i}^{\dag}+E_{j}^{\dag
}\right)  -I\\
&  =\left(  E_{i}E_{i}^{\dag}+E_{j}E_{j}^{\dag}\right)  +\left(  E_{i}%
E_{j}^{\dag}+E_{j}E_{i}^{\dag}\right)  +\left(  E_{i}+E_{i}^{\dag}\right)
+\left(  E_{j}+E_{j}^{\dag}\right) \\
&  =E_{i}E_{j}^{\dag}+E_{j}E_{i}^{\dag}.
\end{align*}
Here the first line uses unitarity, and the fourth line uses the fact that
$E_{i}+E_{i}^{\dag}=-E_{i}E_{i}^{\dag}$ and $E_{j}+E_{j}^{\dag}=-E_{j}
E_{j}^{\dag}$. \ Lemma \ref{matrixlem}\ now implies that there can be at most
$2N$\ nonzero $E_{i}$'s. \ Hence $U\left(X\right)$\ can depend
nontrivially on at most $2N$\ bits of $X$, and can assume at most $2^{2N}$\
values.
\end{proof}

\section{Open Problems\label{OPEN}}

The most obvious problems left open by this paper are, first, to prove a
classical oracle separation between $\mathsf{QMA}$\ and $\mathsf{QCMA}$, and
second, to prove that the Group Non-Membership problem is in $\mathsf{QCMA}$.
\ We end by listing four other problems.

\begin{enumerate}
\item[(1)] The class $\mathsf{QMA}\left(2\right)$\ is defined similarly to
$\mathsf{QMA}$, except that now there are two quantum provers who
are guaranteed to share no entanglement. \ Is there a quantum oracle
relative to which $\mathsf{QMA}\left(2\right) \neq\mathsf{QMA}$?

\item[(2)] Is there a quantum oracle relative to which $\mathsf{BQP/qpoly}
\not \subset \mathsf{QMA/poly}$? \ This would show that the containment
$\mathsf{BQP/qpoly}\subseteq\mathsf{PP/poly}$\ proved in \cite{aar:adv}\ is in
some sense close to optimal.

\item[(3)] Can we use the ideas of Section \ref{SQO}\ to give a classical
oracle relative to which $\mathsf{BQP}\not \subset \mathsf{PH}$? \ What about
a classical oracle relative to which $\mathsf{NP}\subseteq\mathsf{BQP}$\ but
$\mathsf{PH}\not \subset \mathsf{BQP}$?\footnote{Note that a simple
relativizing argument shows that if $\mathsf{NP}\subseteq\mathsf{BPP}$\ then
$\mathsf{PH}\subseteq\mathsf{BPP}$.}

\item[(4)] Is there a polynomial-time quantum oracle algorithm $Q$, such that
for every $n$-qubit unitary transformation $U$, there exists a classical
oracle $A$\ such that $Q^{A}$\ approximately implements $U$? \ Alternatively,
would any such algorithm require more than $\operatorname*{poly}\left(
n\right)  $\ queries to $A$?\footnote{We do not even know whether a
\textit{single} query suffices. \ Note that Theorem \ref{onequery}\ does not
apply here, since we have dropped the requirement that $Q^{A}$\ must implement
\textit{some} $n$-qubit unitary (as opposed to a more general superoperator)
for every oracle $A$.}
\end{enumerate}

\section{Acknowledgments}

We thank the anonymous reviewers for their suggestions, and Dorit Aharonov,
Laci Babai, Robert Beals, Robert Guralnick, Bill Kantor, and Cris Moore for
helpful correspondence.


\begin{thebibliography}{10}

\bibitem{aar:copy}
S.~Aaronson, \emph{Quantum copy-protection}, In preparation.

\bibitem{aar:adv}
\bysame, \emph{Limitations of quantum advice and one-way communication}, Theory
  of Computing \textbf{1} (2005), 1--28, quant-ph/0402095. Conference version
  in Proceedings of CCC'2004.

\bibitem{an}
D.~Aharonov and T.~Naveh, \emph{Quantum {NP} - a survey}, quant-ph/0210077,
  2002.

\bibitem{babai:am2}
L.~Babai, \emph{Trading group theory for randomness}, Proc. ACM STOC, 1985,
  pp.~421--429.

\bibitem{babai:am}
\bysame, \emph{Bounded round interactive proofs in finite groups}, SIAM J.
  Discrete Math \textbf{5} (1992), no.~1, 88--111.

\bibitem{be}
L.~Babai and P.~Erd\H{o}s, \emph{Representation of group elements as short
  products}, Annals of Discrete Math. \textbf{12} (1982), 27--30.

\bibitem{bgklp}
L.~Babai, A.~J. Goodman, W.~M. Kantor, E.~M. Luks, and P.~P. P\'{a}lfy,
  \emph{Short presentations for finite groups}, J. Algebra \textbf{194} (1997),
  no.~1, 79--112.

\bibitem{bs:matrix}
L.~Babai and E.~Szemer\'{e}di, \emph{On the complexity of matrix group problems
  {I}}, Proc. IEEE FOCS, 1984, pp.~229--240.

\bibitem{bclr}
M.~Ben-Or, D.~Coppersmith, M.~Luby, and R.~Rubinfeld, \emph{Non-abelian
  homomorphism testing, and distributions close to their self-convolutions},
  Proceedings of RANDOM, Springer-Verlag, 2004, ECCC TR04-052, pp.~273--285.

\bibitem{bbbv}
C.~Bennett, E.~Bernstein, G.~Brassard, and U.~Vazirani, \emph{Strengths and
  weaknesses of quantum computing}, SIAM J. Comput. \textbf{26} (1997), no.~5,
  1510--1523, quant-ph/9701001.

\bibitem{boroczky}
K.~B\"{o}r\"{o}czky~Jr. and G.~Wintsche, \emph{Covering the sphere by equal
  spherical balls}, Discrete and Computational Geometry: The Goodman-Pollack
  Festschrift, Springer, 2003, pp.~237--253.

\bibitem{bhmt}
G.~Brassard, P.~H{\o}yer, M.~Mosca, and A.~Tapp, \emph{Quantum amplitude
  amplification and estimation}, Quantum Computation and Information (S.~J.
  Lomonaco and H.~E. Brandt, eds.), Contemporary Mathematics Series, AMS, 2002,
  quant-ph/0005055.

\bibitem{atlas}
J.~H. Conway, R.~T. Curtis, S.~P. Norton, R.~A. Parker, and R.~A. Wilson,
  \emph{Atlas of finite groups}, Clarendon Press, Oxford, 1985.

\bibitem{ehk}
M.~Ettinger, P.~H{\o}yer, and E.~Knill, \emph{The quantum query complexity of
  the hidden subgroup problem is polynomial}, Inform. Proc. Lett. \textbf{91}
  (2004), no.~1, 43--48, quant-ph/0401083.

\bibitem{grover:framework}
L.~K. Grover, \emph{A framework for fast quantum mechanical algorithms}, Proc.
  ACM STOC, 1998, quant-ph/9711043, pp.~53--62.

\bibitem{hrt}
S.~Hallgren, A.~Russell, and A.~Ta-Shma, \emph{The hidden subgroup problem and
  quantum computation using group representations}, SIAM J. Comput. \textbf{32}
  (2003), no.~4, 916--934, Conference version in STOC'2000, p. 627-635.

\bibitem{hofling}
B.~H\"{o}fling, \emph{Efficient multiplication algorithms for finite polycyclic
  groups}, 2007, Submitted.
  www-public.tu-bs.de/~bhoeflin/preprints/collect.pdf.

\bibitem{hulpke-seress}
Alexander Hulpke and \'{A}kos Seress, \emph{Short presentations for
  three-dimensional unitary groups}, J. Algebra \textbf{245} (2001), no.~2,
  719--729.

\bibitem{kkr}
J.~Kempe, A.~Kitaev, and O.~Regev, \emph{The complexity of the {L}ocal
  {H}amiltonian problem}, SIAM J. Comput. \textbf{35} (2006), no.~5,
  1070--1097, quant-ph/0406180.

\bibitem{kitaev:meas}
A.~Kitaev, \emph{Quantum measurements and the abelian stabilizer problem}, ECCC
  TR96-003, quant-ph/9511026, 1996.

\bibitem{kitaev:ec}
\bysame, \emph{Quantum computation: algorithms and error correction}, Russian
  Math. Surveys \textbf{52} (1997), no.~6, 1191--1249.

\bibitem{mw}
C.~Marriott and J.~Watrous, \emph{Quantum {A}rthur-{M}erlin games},
  Computational Complexity \textbf{14} (2005), no.~2, 122--152.

\bibitem{mrr}
C.~Moore, D.~N. Rockmore, and A.~Russell, \emph{Generic quantum {F}ourier
  transforms}, Proc. ACM-SIAM Symp. on Discrete Algorithms (SODA), 2004,
  quant-ph/0304064, pp.~778--787.

\bibitem{moscastebila}
M.~Mosca and D.~Stebila, \emph{Unforgeable quantum money}, In preparation,
  2006.

\bibitem{neumann}
P.~M. Neumann, \emph{An enumeration theorem for finite groups}, Quart. J. Math.
  Ser. \textbf{2} (1969), no.~20, 395--401.

\bibitem{pyber}
L.~Pyber, \emph{Enumerating finite groups of given order}, Annals of
  Mathematics \textbf{137} (1993), 203--220.

\bibitem{razshpilka}
R.~Raz and A.~Shpilka, \emph{On the power of quantum proofs}, Proc. IEEE
  Conference on Computational Complexity, 2004, pp.~260--274.

\bibitem{shamir}
A.~Shamir, \emph{{IP=PSPACE}}, J. ACM \textbf{39} (1992), no.~4, 869--877.

\bibitem{shor}
P.~Shor, \emph{Polynomial-time algorithms for prime factorization and discrete
  logarithms on a quantum computer}, SIAM J. Comput. \textbf{26} (1997), no.~5,
  1484--1509, Earlier version in IEEE FOCS 1994. quant-ph/9508027.

\bibitem{sims}
C.~Sims, \emph{Computational methods in the study of permutation groups},
  Computational Problems in Abstract Algebra, Pergamon Press, 1970,
  pp.~169--183.

\bibitem{watrous}
J.~Watrous, \emph{Succinct quantum proofs for properties of finite groups},
  Proc. IEEE FOCS, 2000, cs.CC/0009002, pp.~537--546.

\end{thebibliography}

\providecommand{\bysame}{\leavevmode\hbox to3em{\hrulefill}\thinspace}
\providecommand{\MR}{\relax\ifhmode\unskip\space\fi MR }
\providecommand{\MRhref}[2]{%
  \href{http://www.ams.org/mathscinet-getitem?mr=#1}{#2}
}
\providecommand{\href}[2]{#2}
\providecommand{\eprint}{\begingroup \urlstyle{tt}\Url}

\end{document}